\documentclass[10pt,conference,letterpaper]{IEEEtran}

\usepackage{amsmath,amssymb,amsfonts}
\usepackage{algorithmic}
\usepackage{cite}
\usepackage{hyperref}
\usepackage{cleveref} % Must be loaded after hyperref
\usepackage{graphicx}
\usepackage{makecell}
\usepackage{siunitx}
\usepackage{subcaption}
\usepackage{textcomp}
\usepackage{tikz}
\usepackage{xcolor}
\def\BibTeX{{\rm B\kern-.05em{\sc i\kern-.025em b}\kern-.08em
    T\kern-.1667em\lower.7ex\hbox{E}\kern-.125emX}}

\newcommand{\iseditmode}{0}
\newcommand{\editmode}{\renewcommand{\iseditmode}{1}}

\newcommand{\addtext}[1]{\if\iseditmode1{{\leavevmode\color{blue}#1}}\else{#1}\fi}
\newcommand{\replacetext}[2]{\if\iseditmode1{{\leavevmode\color{red}#1} {\leavevmode\color{blue}#2}}\else{#2}\fi}
\newcommand{\removetext}[1]{\if\iseditmode1{{\leavevmode\color{red}#1}}\fi}

\DeclareMathOperator{\sgn}{sgn}

\DeclareMathOperator{\GCL}{GCL}
\DeclareMathOperator{\RTT}{RTT}

\DeclareMathOperator{\nxt}{nxt}

\newcommand*\delimeter[3]{\ensuremath{\mathopen{}\left#2 #1\right#3\mathclose{}{\vphantom{\left#2 #1\right#3}}}}

\newcommand*\pof[1]{\delimeter{#1}{(}{)}}

\newcommand*\cof[1]{\delimeter{#1}{\{}{\}}}

\newcommand*\abs[1]{\delimeter{#1}{|}{|}}
\newcommand*\norm[1]{\delimeter{#1}{\|}{\|}}

\newcommand*\dif{\ensuremath{\mathrm{d}}}

\editmode
\crefname{figure}{Fig.}{Figs.}
\Crefname{figure}{Fig.}{Figs.}
\crefname{section}{section}{sections}
\Crefname{section}{Section}{Sections}
\crefname{table}{Table}{Tables}
\Crefname{table}{Table}{Tables}

\begin{document}

\title{Matisse: Visualizing Measured Internet Latencies as Manifolds}

\author{\IEEEauthorblockN{1\textsuperscript{st} Stephen Jasina}
\IEEEauthorblockA{\textit{University of Wisconsin-Madison} \\
%Madison, Wisconsin \\
jasina@wisc.edu}
\and
\IEEEauthorblockN{2\textsuperscript{nd} Loqman Salamatian}
\IEEEauthorblockA{\textit{Columbia University} \\
%New York, New York \\
ls3748@columbia.edu}
\and
\IEEEauthorblockN{3\textsuperscript{rd} Joshua Mathews}
\IEEEauthorblockA{\textit{Carnegie Mellon University} \\
%Pittsburgh, Pennsylvania \\
jcmathew@andrew.cmu.edu}
\and
\IEEEauthorblockN{4\textsuperscript{th} Scott Anderson}
\IEEEauthorblockA{\textit{University of Wisconsin-Madison} \\
%Madison, Wisconsin \\
standerson4@wisc.edu}
\and
\IEEEauthorblockN{5\textsuperscript{th} Paul Barford}
\IEEEauthorblockA{\textit{University of Wisconsin-Madison} \\
%Madison, Wisconsin \\
pb@wisc.edu}
\and
\IEEEauthorblockN{6\textsuperscript{th} Mark Crovella}
\IEEEauthorblockA{\textit{Boston University} \\
%Boston, Massachusetts \\
crovella@bu.edu}
\and
\IEEEauthorblockN{7\textsuperscript{th} Walter Willinger}
\IEEEauthorblockA{\textit{NIKSUN Inc.} \\
%Princeton, New Jersey \\
wwillinger@niksun.com}
}

\maketitle

\begin{abstract}
Manifolds are complex topological spaces that can be used to represent datasets of real-world measurements. Visualizing such manifolds can help with illustrating their topological characteristics (\emph{e.g.}, curvature) and providing insights into important properties of the underlying data (\emph{e.g.}, anomalies in the measurements). In this paper, we describe a new methodology and system for generating and visualizing manifolds that are inferred from actual Internet latency measurements between different cities and are projected over a 2D Euclidean space (\emph{e.g.}, a geographic map). Our method leverages a series of graphs that capture critical information contained in the data, including well-defined locations (for vertices) and Ricci curvature information (for edges). Our visualization approach then generates a curved surface (manifold) in which (a) geographical locations of vertices are maintained and (b) the Ricci curvature values of the graph edges determine the curvature properties of the manifold. The resulting manifold highlights areas of critical connectivity and defines an instance of ``Internet delay space" where latency measurements manifest as geodesics. We describe details of our method and its implementation in a tool, which we call {\em Matisse}, for generating, visualizing and manipulating manifolds projected onto a base map. We illustrate Matisse with two case studies: a simple example to demonstrate key concepts, and visualizations of the US public Internet to show Matisse's utility.
\end{abstract}


\section{Introduction}             \label{sec:introduction}  In studying communication networks such as the Internet, one frequently encounters datasets of measurements representing the network as a graph, with various kinds of node and edge attributes. An important example is graphs having nodes embedded in a (2D) Euclidean space ({\em i.e.,} with latitude and longitude) and having edge attributes in the form of measured latencies ({\em i.e.,} delays incurred by packets traveling between pairs of nodes at a particular instant).  Interpreting such simultaneously collected per-edge measurement datasets is important for understanding the behavior of the complex underlying collection of cables, routers, and data centers that comprise the Internet.

\begin{figure}[ht]
    \centering
    \includegraphics[width=0.8\columnwidth]{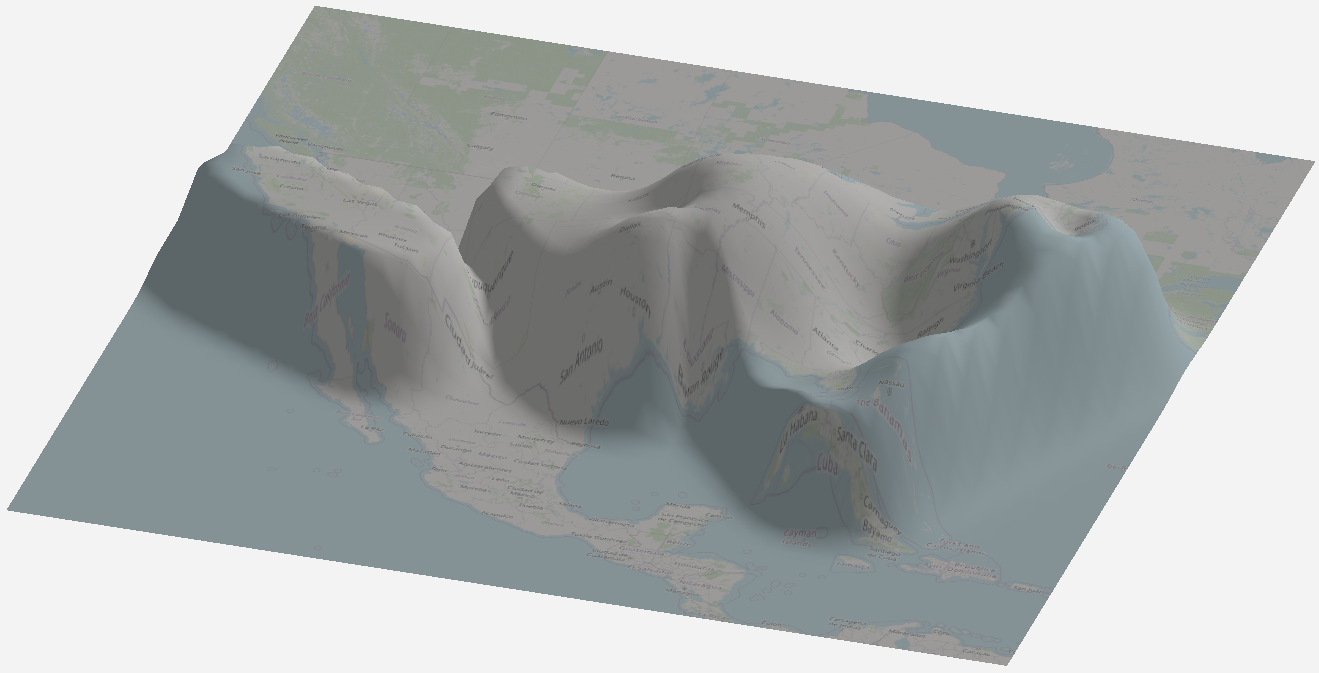}
    \caption{A Matisse-based visualization of an instance of Internet delay space (US public Internet). Higher areas in the northeast indicate denser connectivity with low delays (\emph{e.g.}, geodesics are close to line-of-sight), while the valley that bifurcates the country indicates an area of sparse but critical connectivity between the eastern and western parts of the US causing higher delays (\emph{e.g.}, geodesics are far from line-of-sight).}
    \label{fig:public_us}
\end{figure}

Our present work is based on the hypothesis that such datasets allow for geometric representations as complex metric spaces in the form of smooth manifolds embedded in 3D. In the case of Internet delay measurements, we refer to such a manifold representation as ``Internet delay space'' and show in this paper that it offers new opportunities for both developing a new visualization technique, showcased in \cref{fig:public_us}, and characterizing different aspects of observed Internet latency measurements~\cite{sigmetrics2022, cacm_23}.

A manifold representation of measured Internet latencies is informative because of its contrast with a flat Euclidean space.  If we treat geo-located Internet nodes as approximately lying on a flat 2D (Euclidean) surface in physical space, then a manifold can depict the type and amount of deformations of this flat surface that are required to capture measured latencies. That is, we can construct a visualization such that the shortest path between any two nodes in the manifold approximately reflects the measured latency between the two nodes in question. Then, by ``walking along the manifold'',  the measured Internet latencies between different pairs of nodes can be faithfully recovered from the geodesic length between the corresponding pairs of nodes. For example, while low latency measurements often result from paths that correspond to notions such as ``line-of-sight'' or ``walking on a locally Euclidean plane,'' high-latency measurements tend to be associated with less-direct paths. In the manifold view of the resulting Internet delay space, such paths have to traverse ``hills and valleys'' that are caused by deformations of the Euclidean plane and manifest as curved surfaces.

Hence, the goal of our work is to develop a visualization that:
{\em (a)} is computed from known node locations and measured inter-node latencies;
{\em (b)} is presented as a manifold with a clear link to real-world geography;
{\em (c)} enables macroscopic interpretation of network properties while accurately reflecting latency relationships between individual nodes through distance on the surface; and
{\em (d)} handles simultaneous delay measurements between multiple node pairs, thus capturing the core structure of the underlying delay space in an interpretable way.

To achieve this goal in practice, we face the challenge that Internet latencies arise in a space that is hidden from us. It is hidden because the underlying physical infrastructure that individual packets traverse from one node to another is generally only partially known. Furthermore, the geography of the physical space in which the nodes reside influences the structure of measured latencies, and the precise relationship between the physical and latency-defined spaces is not known.

However, even though end-to-end packet latencies in the Internet are not necessarily symmetric and may not even respect the triangle inequality, we can take steps to address these issues ({\em e.g.,} by working with packet round-trip time measurements (RTTs) as basic entities). Since the scale of these modifications is small compared to the magnitudes of the measurements we use and their impact on our visualizations is minimal, we can treat this hidden space for all practical purposes as a metric space.

Our approach to overcoming the challenges posed by measured Internet latencies relies on examining the {\em differences} between the delays in the hidden space ({\em i.e.,} the measured Internet latencies) and the delays in the physical space ({\em i.e.,} based on geographic distance on the map).  We argue that understanding these differences provides insight into the nature of the network measurements and the properties of the underlying network infrastructure. The principal tool we use to interpret these differences and extract insight is the \emph{Ollivier-Ricci} curvature (hereafter shortened to Ricci curvature).

Hence, our strategy is to leverage Ricci curvature to construct a visualization that \emph{both} reflects the hidden delay space, \emph{as well as} the position of nodes in the known map space.  We reconcile these two objectives by using a 2D manifold embedded in 3D space and adapting the local manifold curvature to match the curvature implied in each region of the delay measurements.  The resulting manifold is a reflection of the deformation of the Euclidean space that is imposed by the hidden delay space; ideally, the manifold's geodesics approximately encode the measured Internet latencies.

Our main contribution is \emph{Matisse}, a visualization system designed to meet the criteria laid out above.  Matisse relies on {\em (a)} creating graphs derived from both the hidden and underlying map spaces; {\em (b)} characterizing the resulting graphs using Ricci curvature; and {\em (c)} using the derived Ricci curvature values to determine the manifold's deformation.

To illustrate Matisse's functionality, flexibility, and utility, we present two case studies.  The first case study utilizes a toy network that omits a specific geographic mapping of nodes and is presented to highlight the basic aspects of manifold generation and visualization. The second case study is based on packet delay measurements between a set of hosts with known locations in the US public Internet.  The obtained manifold is a representation of the inferred Internet delay space and highlights areas of interest within that topological space.  Finally, we use the manifolds generated for measured delays in the Internet to demonstrate that our method produces a surface that accurately represents the hidden metric space.  We do this by showing that the geodesic distances between vertices estimated on the manifold's surface are a more accurate reflection of the measured delays than the line-of-site distance.  The curved nature of the resulting manifold thus encodes the measured Internet latency between any two nodes as shortest paths between them on the surface of the manifold.

\section{Related Work}             \label{sec:background}    Prior studies that inform our work fall into two categories: (1) manifold visualization methods, and (2) geographic mapping techniques that incorporate network or performance data.

\subsection{Learning and Visualizing Manifolds}

\subsubsection{Discovering Manifolds}: Many real-world datasets are high-dimensional but tend to lie on lower-dimensional manifolds. Classical dimensionality reduction techniques such as Principal Component Analysis (PCA)\cite{wold1987principal}, t-SNE\cite{vandermaaten08}, and UMAP~\cite{mcinnis20} attempt to uncover low-dimensional representations of such data for the purpose of visualization or learning.

While these techniques are well-suited for abstract feature spaces, our setting differs in key ways. We work with network measurement data that already has concrete geographic coordinates and measured latency values between nodes. Our goal is to produce interpretable maps that reflect both geography and network performance, which standard methods like PCA or t-SNE are not designed to preserve. In contrast to general-purpose projection tools, we treat latency measurements as defining a hidden metric space and use curvature to reconcile it with the physical space of node locations.

\subsubsection{Curvature} Ricci curvature provides a geometric abstraction for comparing metric properties of neighboring regions. In our work, we use Ollivier-Ricci curvature~\cite{ollivier2009ricci} to characterize how measured latencies between nodes deviate from Euclidean expectations, enabling us to represent these deviations as deformations of a 2D surface embedded in 3D. Prior work has applied Ricci curvature to Internet connectivity graphs. Ni et al.~\cite{Ni15} used curvature to study connectivity within AS-level topologies, showing that geometric properties reveal weakly connected regions. While our use of curvature is inspired by these ideas, we focus on latency measurements and embed them in a continuous manifold tied to geographic space, rather than abstract graph structures.

In prior work~\cite{sigmetrics2022,cacm_23}, we describe a method for  exposing and visualize critical aspects of Internet connectivity using latency measurements and Ricci curvature.  This paper builds on that work by expanding and generalizing the process of manifold generation and delivering a tool -- Matisse -- to the community, which can be applied to publicly available latency datasets.

\subsection{Drawing Augmented Maps}
Traditional cartograms~\cite{sun2010effectiveness, cauvin1989cartographic} distort geographic regions to reflect external quantities like population or GDP. Our setting is an extension of this idea, where we aim to encode measured network performance (latency) into the surface geometry itself, rather than via regional area distortion. This introduces a fundamental tradeoff: maintaining a faithful representation of geography while also accurately reflecting network topology and performance.

Matisse addresses this challenge by deforming the 2D surface using curvature derived from latency measurements. The result is a manifold visualization that retains geographic context while revealing structure in the hidden delay space.

\section{Methodology}              \label{sec:methods}       In this section, we present our methodology for generating manifolds of hidden metric spaces, and we describe the Matisse system, which implements our methodology.  While our method and visualization system are general in the sense that they can be applied to graphs projected onto any Euclidean surface, for the remainder of this section, we will concentrate on graphs projected onto \emph{geographic surfaces}, as these are both intuitive and a particularly relevant application for our technique.

\subsection{Overview}
\label{sec:basics}

Our first key observation is that the Internet's delay space resembles a spatial map, where each connected device acts as a point. These points are positioned based on their physical locations and the network delay, or latency, between them. Round trip latency measures how long it takes for data to travel from one point to another and back, and is affected by factors like network congestion, routing practices, and infrastructure availability. In regions lacking infrastructure, connectivity becomes stretched, much like physical terrain can develop gaps or folds due to erosion. This phenomenon creates a particular effect in our representation: it causes the "space" of Internet latencies to bend or curve. This curvature, reminiscent of the way general relativity describes the shape of spacetime, offers an intuitive way to model the distortion between the physical and latency space.

Ricci curvature measures, at each point in space, how much space diverges from being flat. In a continuous setting, this involves using tensors to detail the curvature at every point, describing the space's shape at all points---whether it gathers like a valley, extends flat like a plane, or twists like a saddle. The main idea behind Matisse is to leverage this analogy to curvature and manifolds to provide a powerful method for viewing and analyzing the latency space.

Bridging the gap between continuous and discrete settings is crucial, specifically when trying to represent the continuous geometric features of the Internet using discrete measurements between vantage points. Our solution utilizes a novel technique to infer a discrete curvature from these vantage points~\cite{sigmetrics2022} (described in \cref{sec:input_output,sec:threshold,sec:computing_ricci}) and the use of meshes to bridge the gap between discrete and continuous. By fine-tuning the mesh's level of detail (granularity), we can more accurately represent the continuous aspects of curvature. When dealing with 2D surfaces in 3D space, we replace the Ricci Curvature with its equivalent Gaussian curvature, whose properties are more computationally tractable. This approach not only enables us to depict the Internet's structure through its geometric characteristics but also offers a method to assess and analyze the network in a manner that is consistent with our understanding of space and curvature in a continuous world. In this section, we describe how we generate our figures, adopting a bottom-up approach from raw latency measurements to the final output.

\subsection{Inputs and Outputs}
\label{sec:input_output}
Matisse takes as input a list of geo-located (latitude and longitude) vantage points and a set of measured round trip times between them. Together, these form a (nearly, in the case of possibly missing measurements) complete weighted graph. Additional inputs to Matisse are the tuning parameters $\varepsilon$, $\lambda_{\text{smooth}}$, $r$; an initialization; a clustering distance; and the number of vertices to use in a mesh. The selection of these hyperparameters is discussed in \cref{sec:parameters}.

Matisse produces a weighted simple graph $G = (V, E)$ as an intermediate output. $G$'s vertices represent the vantage points, and its edges are determined by a thresholding process. The weight of each edge $e \in E$ is its Ricci curvature $\kappa_{\text{Ricci}}(e)$. The thresholding process and the Ricci curvature computation are both described below.

As output, Matisse returns a triangle mesh representing a manifold that reflects the values of $\kappa_{\text{Ricci}}$ and in which each vertex $v \in V$ has a corresponding location $\pi(v)$.

\subsection{Thresholding Process}
\label{sec:threshold}

An important application of Matisse is in analyzing differing network representations of a set of measured latencies. This goal requires an analysis process that differs from traditional techniques that reduce high-dimension data into a single visualization ({\em e.g.,} via PCA). Our analysis process allows for generation of manifolds based on the threshold value $\varepsilon$, defined as the difference (residual) between an idealized ({\em e.g.,} straight line) distance between two vantage points and the measured metric ({\em e.g.,} RTT) between those nodes. For the given value of $\varepsilon$, we construct edges between nodes whose residual values are at most $\varepsilon$. Thus, different thresholds can result in different graphs that can be input to Matisse.

Our definition of residuals implies that smaller residuals correspond to more direct network paths, as only fairly direct paths have latencies that approach the idealized values. Selecting a relatively small (less than $25$ ms) $\varepsilon$ should therefore produce a graph that is representative of the actual network connectivity. By progressing from lower to higher values for $\varepsilon$, we generate manifolds that can include key features that emerge from the use of Ricci curvature. These include {\em saddles} on negatively curved edges and {\em mountains} or \emph{valleys} in areas of positively curved edges. It is important to emphasize that we are particularly interested in areas of negative curvature in the resulting manifolds since these relate to edges that serve as critical connectors between regions of positive curvature. Examples of these features will be shown below and in the case studies described in \cref{sec:results}.

When the thresholding process is complete, we have an unweighted version of $G$.

\subsection{Computing Discrete Ricci Curvature}
\label{sec:computing_ricci}

Ricci curvature fundamentally draws upon the parallels  between optimal transport theory -- which considers transportation of resources from one location to another in the most efficient manner possible -- and the notion of curvature. The relationship hinges on the idea that curvature affects the cost of transporting mass in space; in areas of different curvature, the optimal paths and costs for transportation vary, reflecting the underlying geometry's space. Formally, the transportation distance, denoted as $W_1(m_x,m_y)$, measures the "cost" to transform one probability distribution $m_x$ into another $m_y$ within a given metric space $(X,d)$. The transportation distance is calculated as the smallest possible total distance needed to move "mass" from the distribution $m_x$ to match the distribution $m_y$. 
\begin{equation}
    W_1(m_x, m_y) := \inf_{\eta \in \prod(m_x,m_y)} \int_{X \times X} d(x,y) d\eta(x,y)
\end{equation}
Here, $\prod(m_x,m_y)$ represents all joint distributions on $X \times X$ that have $m_x$ and $m_y$ as their margins. The Ricci curvature $\kappa(x,y)$ between two neighboring vertices $x$ and $y$ is defined as
\begin{equation}
    \kappa_{\text{Ricci}}(x,y) = 1 - W_1(m_x, m_y),
\end{equation}
where the measure $m_x$ distributes equal mass across all of $x$'s neighbors. Specifically, if $x$ has a degree of $d_x$ (meaning, $x$ is directly connected to $d_x$ other vertices), then each neighbor of $x$ is assigned a weight of $1/d_x$ under the measure $m_x$. As we refine the discretization of a space, making it more closely resemble a continuous domain, the Ricci curvature begins to emulate the behaviors and properties of Ricci curvature in a smooth manifold~\cite{ollivier2009ricci,hoorn2023ollivier}. This convergence property offers a natural framework for analyzing the intrinsic geometric properties of the latency space.

While the definition of Ricci curvature is quite involved, we can develop some intuition by looking at simple graphs. In a tree, the neighbor sets of two adjacent nodes are disjoint, so nearly every piece of mass in an optimal transportation plan would need to move a distance of $3$ ($1$ to get to $x$, $1$ along the $x$-$y$ edge, and $1$ to leave from $y$). Consequently, $\kappa_{\text{Ricci}}$ will be approximately $-2$.

In the other extreme, the neighbor sets of two adjacent nodes in a complete graph nearly coincide, so the optimal transportation cost should be close to $0$. $\kappa_{\text{Ricci}}$ will thus be approximately $1$ for any edge in a complete graph.

Putting these two ideas together, we find that $\kappa_{\text{Ricci}}$ increases as connectivity increases. The curvature is most negative for edges that are critical to the (local) connectedness of the graph.

\subsection{Generating Manifolds}
\label{sec:generation}
\subsubsection{Translating graph curvature to surface curvature} A central question in generating visualizations of manifolds of hidden metric spaces concerns how to translate the discrete curvature on edges in the planar input graph into a continuously curved surface embedded in 3D. While it is known that the discrete Ollivier-Ricci curvature converges to the standard continuous Ricci curvature as the sampling points increase in the neighborhood of a point, we control the curvature of the surface using the Gaussian curvature. Each edge influences a patch of the manifold based on its Ricci curvature. Negative Ricci curvatures induce patches of negative Gaussian curvature, while positive Ricci curvatures induce patches of positive curvature.

The goals of our visualization tool are to {\em (1)} map the graph onto a manifold such that the locations of vertices are maintained and the Ricci curvature along any edge in the graph is reflected in the curvature between the corresponding points on the surface, and {\em (2)} to maintain smoothness across the surface.

\subsubsection{Constructing the surface} We approximate our manifold with a triangle mesh. The mesh vertices have fixed positions in the $xy$-plane and form a $k \times k$ uniformly spaced grid. The height of each vertex is then modified to change the curvature of the mesh. To compute the heights, we minimize the loss function
\begin{equation}
    \mathcal{L}(M) = \mathcal{L}_{\text{curvature}}(M) + \lambda_{\text{smooth}}\mathcal{L}_{\text{smooth}}(M)
\end{equation}
across valid choices of mesh $M$. Here, $\mathcal{L}_{\text{curvature}}$ captures how closely the Gaussian curvature of the mesh matches the Ricci curvature of the network, and $\mathcal{L}_{\text{smooth}}$ measures how smooth the mesh is. $\lambda_{\text{smooth}}$ is a hyperparameter that controls how much to favor each of these aspects when optimizing (higher means we value smoothness more).

\subsubsection{Defining the curvature loss} To establish the curvature loss function, we initially define the area surrounding an edge in the mesh. This is achieved through the projection $\pi$ which maps the vertices $V$ onto $xy$-coordinates in $\mathbb{R}^2$ ({\em e.g.,} using the Mercator projection). An edge $e = (u,v) \in E$ can be represented in $\mathbb{R}^2$ as the line segment that joins $\pi(u)$ and $\pi(v)$. We define $B_r(e)$ to be the ball of radius $r$ around edge $e$, {\em i.e.}, the collection of mesh vertices that are within an $xy$-distance of $r$ or less from $\pi(e)$.

We make use of $B_r(e)$ to define $\mathcal{L}_{\text{curvature}}$: \begin{align}
    \mathcal{L}_{\text{curvature}}(M) &= \frac{1}{\sum_{e = (u, v) \in E}\|\pi(u) - \pi(v)\|_2} \nonumber \\
    &\qquad\sum_{e = (u, v) \in E}\frac{\|\pi(u) - \pi(v)\|_2}{|B_r(e)|} \nonumber \\
    &\qquad\sum_{p \in B_r(e)}(\kappa_{\text{Ricci}}(e) - \kappa_{\text{Gaussian}}(p))^2.
\end{align} The innermost summand measures how closely the Gaussian curvature at a point $p$ close to $e$ matches the Ricci curvature of $e$. Multiplying by $\frac{\|\pi(u) - \pi(v)\|_2}{|B_r(e)|}$ ensures each edge is weighted by its length; dividing by $\sum_{e = (u, v) \in E}\|\pi(u) - \pi(v)\|_2$ ensures the loss has roughly comparable values across different networks.

To bring intuition to this complicated-looking definition, minimizing $\mathcal{L}_{\text{curvature}}(M)$ is the same as ensuring that for each network edge $e$ and for each point $p$ within a distance $r$ from $e$, the mesh's curvature at $p$ is close to the network's curvature at $e$.

\subsubsection{Defining the smoothness loss} To quantify the smoothness of a triangle mesh, we adopt an energy functional from the well-studied field of surface fairing. As curvature is at the center of our representation, we use the curvature-based functional called MVS-cross from \cite{Joshi2007EnergyMF}. Defining $\kappa_1$ and $\kappa_2$ to be vectors of the principal curvatures, $L$ to be the (cotangent) Laplacian of the mesh, and $A$ to be the mesh's total surface area, we have \begin{equation}
    \mathcal{L}_{\text{smooth}}(M) = -(\kappa_1^\intercal L\kappa_1 + \kappa_2^\intercal L\kappa_2) \cdot A.
\end{equation} Intuitively, the use of the Laplacian means $\mathcal{L}_{\text{smooth}}$ measures the ``noisiness'' of the two principal curvatures. The rationale for using this particular functional is that we do not want curvature to vary too severely over the mesh.

\subsubsection{Optimizing the objective function} We pass the loss function into an L-BFGS-B solver (with no constraints) to find $\mathcal{L}$'s minimizer. For ease of implementation, we define $\mathcal{L}$ in Python and use SciPy's \texttt{optimize.minimize} solver~\cite{2020SciPy-NMeth}.

\subsubsection{Preprocessing areas of dense connectivity} Given a set of RTT latencies, it is not always desirable to include every node representing a vantage point in the graph. A large cluster of nearby edges can be difficult to interpret in terms of $\mathcal{L}_{\text{curvature}}$, especially if some of the edges have positive curvature and others have negative curvature. Also, potentially undesirable are geographically miniscule connected components, as they distract from more important features such as long edges with negative curvature. Finally, nodes that are very close to each other might lie in the same mesh face if the mesh's size is too small, meaning the computed manifold might not accurately reflect the input graph.

To address these issues, we set a lower bound on how far apart two nodes must be to be considered distinct. To enforce this limit, we apply single-linkage agglomerative hierarchical clustering to the set of nodes. If we stop the clustering once a certain distance threshold is met ({\em e.g.,} $500$ kilometers), then we have ensured that any pair of resulting clusters is at least that distance apart.

Once we have the clusters, we simply select a single node from each cluster for the vertices. The latency between two clusters is defined as the minimum latency between any pair of nodes in each of the clusters. We can thus think of the overall clustering process as collapsing nearby nodes together.

\subsubsection{Post-processing the output} Though the output of the above process is optimized with respect to our objective function, it is sometimes not well-suited for visualization purposes. The main issues are that the overall surface might be curved ({\em e.g.,} a cylinder with some bumps on it) and that the surface might have features outside of the projection of the network ({\em e.g.,} hills that might decrease the smoothness loss but distract from the pertinent features).

To address the first issue, we can simply subtract the initial heights from the final heights. As long as the principal curvatures of the initial mesh are close to $0$ and the initial mesh is relatively flat (as is the case for our sphere cap), the curvatures of the resulting surface will not be changed significantly. Although this strategy is not guaranteed to help, some experimentation with several networks reveals that it often produces manifolds that are more interpretable.

The second issue requires us to decide what it means to be inside the projection of the network. There are a few choices for this, but one that works well is to use the union of the convex hulls of each of the network's connected components. We can then flatten the surface by scaling the height at a point according to the distance to the network interior. The way that we do this is to shift the entire surface so it lies above the $xy$-plane and then set the height of all points a sufficient distance away from the network interior to $0$.

\subsection{Parameter Choices and Practical Considerations}
\label{sec:parameters}
The optimization relies on a number of user-supplied hyperparameters. Their values need to be carefully selected so as to emphasize the desired properties of the resulting manifold. We provide some heuristics to guide their selection.

$\varepsilon$ controls which latency measurements are used in constructing the manifold. Smaller values will highlight more local structure. In the case study in \cref{sec:internet_US,sec:manifold_distance}, $\varepsilon$ ranges from $10 \unit{\milli\second}$ to $22 \unit{\milli\second}$.

$\lambda_{\text{smooth}}$ controls how jagged the resulting manifold can be. A smaller value places a higher importance on matching the curvatures. We find that values between $0.0002$ and $0.002$ typically produce visually appealing output. In this paper, we optimize over several values of $\lambda_{\text{smooth}}$ and choose a visually appealing output.

$r$ controls the width of the balls around edges. Because we want points in the tube to be close to the line segment, we choose $r$ to be rather small. In particular, we select $r$ to be the length of the longest edge in the flattened mesh. This choice ensures no part of the tube is ``one point wide.''

Some experimentation reveals that initializing the mesh to be a sphere cap tends to produce smooth meshes where areas of positive curvature are mountains, as opposed to valleys.

For clustering, we choose a distance threshold that exploits domain knowledge. More precisely, for our real world examples, we select $500 \unit{\kilo\meter}$ so that nodes in the same metro area are clustered together.

Finally, the number of mesh vertices affects how detailed the resulting manifold can be. We find that a $50 \times 50$ mesh always suffices for our examples, and sometimes smaller meshes provide equally good output.

\subsection{Implementation}
\label{sec:implementation}

The Matisse application front end is written in HTML, CSS, and Javascript. We use the three.js graphics library~\cite{threejs} to build the visualization based on a triangular mesh. We overlay a world map on the plane as a texture using OpenLayers~\cite{openlayers} to get the map data. The back end is entirely written in Python. We make great use of the standard NetworkX, NumPy, scikit-learn, and SciPy libraries for general computation. We additionally use the POT~\cite{flamary2021pot} library for Ricci curvature computation.  The Matisse codebase is publicly available on GitHub\cite{matisseRepository}.

\section{Matisse in Action}        \label{sec:results}       We illustrate Matisse and the applicability of our manifold visualization technique with two case studies.  The first uses simple graphs to highlight the key aspects of our manifold generation and refinement process. The second case study concerns manifold representations of sets of Internet latency measurements between locations in the US public Internet. Similar efforts that focus on the private backbone infrastructures of three large cloud provider networks are described in~\cite{sigmetrics2022}.

\subsection{Toy Network}

\begin{figure}[!ht]
    \centering
    \begin{subfigure}{\columnwidth}
        \centering\includegraphics[width=0.4\columnwidth]{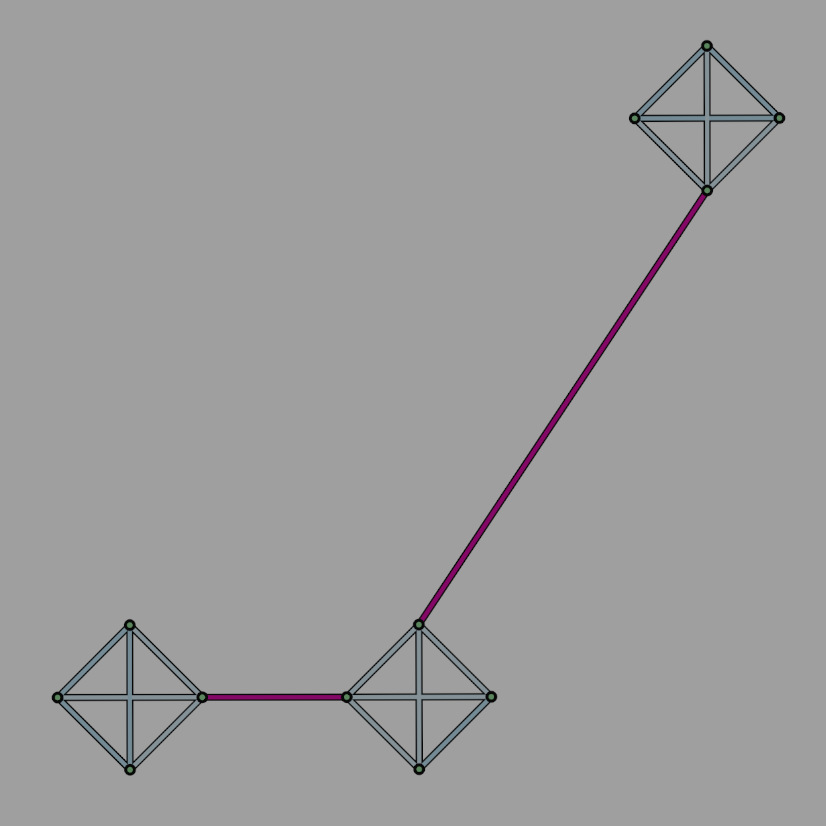}
        \caption{The network itself. Blue links have positive curvature; red links have negative curvature.}
        \label{fig:toy_graph}
    \end{subfigure}
    \begin{subfigure}{\columnwidth}
        \centering\includegraphics[width=0.8\columnwidth]{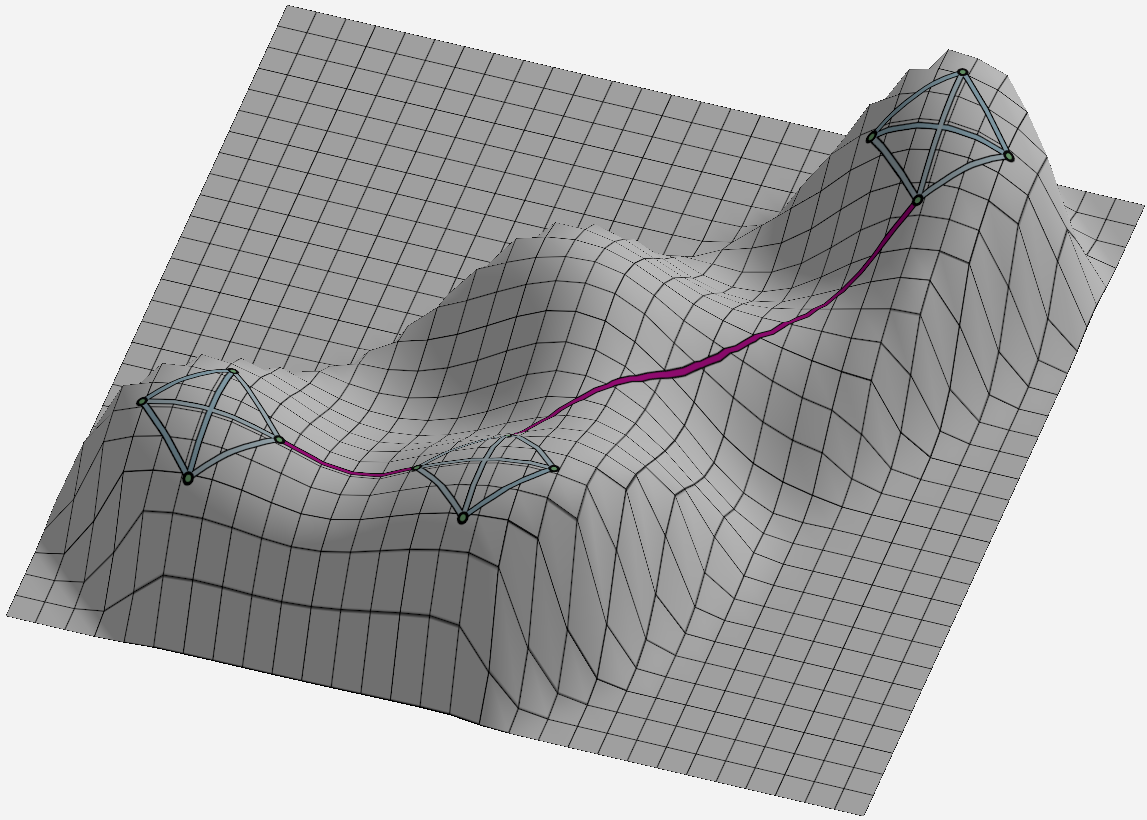}
        \caption{Toy manifold}
        \label{fig:toy_manifold}
    \end{subfigure}
    \begin{subfigure}{\columnwidth}
        \centering\includegraphics[width=0.8\columnwidth]{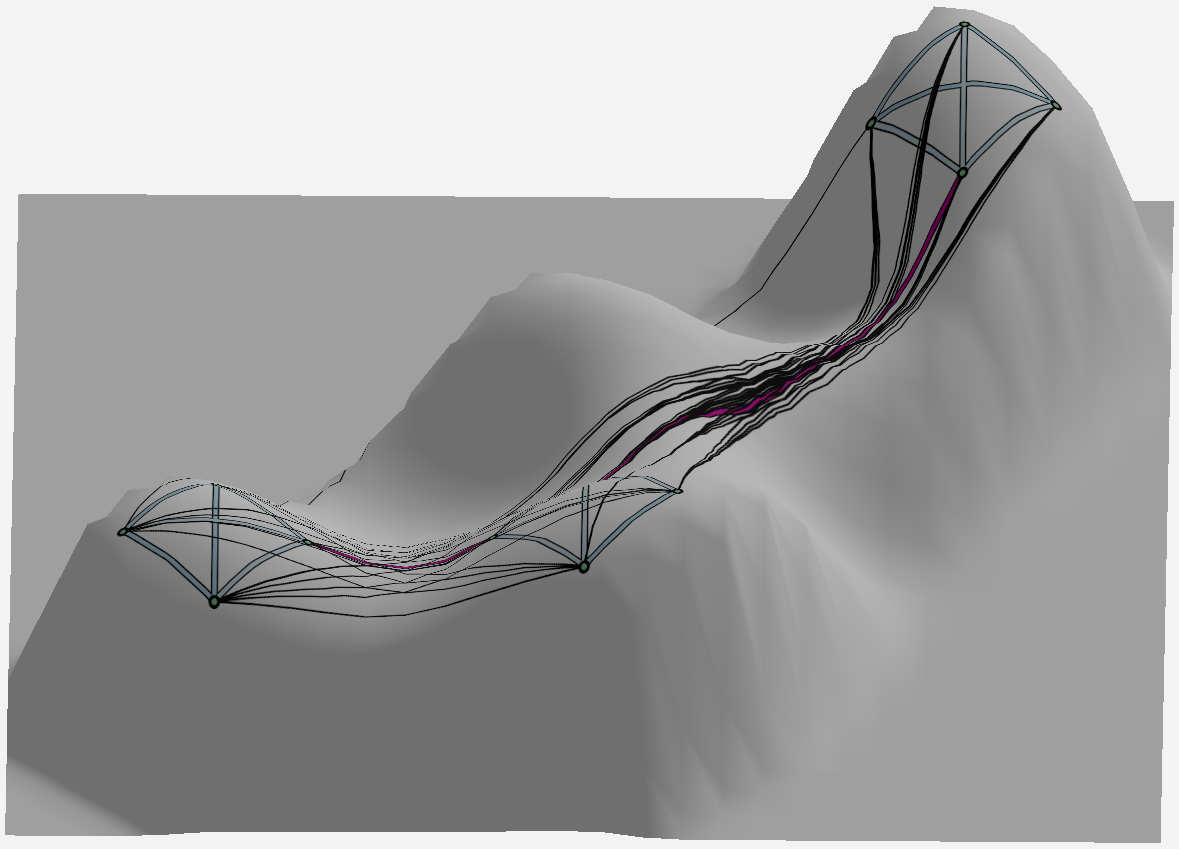}
        \caption{Toy manifold with geodesic paths drawn in black. Note attraction of the paths to the saddles}
        \label{fig:toy_geodesics}
    \end{subfigure}
    \caption{Toy network with its resulting manifold.}
    \label{fig:toy}
\end{figure}

Our first case study is based on the simple graph shown in \cref{fig:toy_graph}. The graph has three regions of ``dense'' connectivity indicated by edges with positive curvature (blue). The regions are connected by links with negative curvature (red). The graph is loaded into Matisse with node locations on a simple Euclidean plane.

Matisse generates the manifold for the graph shown in \cref{fig:toy_manifold}, where the graph is projected down onto the surface.

\Cref{fig:toy_geodesics} shows the effect that imposing curvature has on the geodesics. As our intuition suggests, geodesic paths are attracted to the areas of negative curvature. These negative curvature edges are the ones that are critical to network connectivity. In other words, the geodesic paths approximately match the shortest paths found in the network.

Simple examples like this are powerful because they allow us to investigate the effects of changing the hyperparameters on the features of the resulting manifold. For example, experimenting with the grid size reveals that a mesh that is too coarse loses desirable features like prominent saddles, and increasing the refinement can have diminishing returns. Of course, increasing the refinement incurs an increase in the computation time. In practice, selecting the coarsest granularity that still shows the desired features typically requires experimenting with a range of different grid sizes.

\subsection{Measured Delay Space for the US Public Internet}
\label{sec:internet_US}

The connectivity structure of the Internet has been the subject of intensive study for many years ({\em e.g.,} see \cite{willinger2013, huffaker2002} and references therein).  One way to consider Internet connectivity is to construct a graph based on Round Trip Time packet latency measurements between a full mesh of vantage points distributed throughout the network using the standard ping tool~\cite{ping}.  This simple perspective can provide useful insights into Internet connectivity that can be highlighted in a manifold visualization.

To generate visualizations of manifolds of Internet delay spaces in Matisse, we organize data as follows.  First, we utilize the RIPE Atlas infrastructure~\cite{ripeatlas} to gather RTT measurements from a selected set of 21 vantage points in the continental US. These vantage points are geographically distributed throughout the major metropolitan areas in the US and they are physically connected to each other through long haul fiber conduits that generally follow rights-of-way that mirror major highway and rail infrastructure~\cite{rama-map}.  Important to Matisse, the geographic locations of all vantage points are known.  This allows us to calculate the {\em great circle distance} (GCD) between vantage points.  We then calculate {\em great circle latency} (GCL) between two locations by dividing GCD by $(2/3)c$, which is the speed of light through optical fiber.  We then compute {\em residual latency} for each pair of vantage points, which we define to be the minimum latency (measured over some fixed period of time) minus the GCL.

We generate visualizations of this Internet delay space in Matisse by entering node locations and Ricci curvature values.  We determine connectivity between nodes by selecting edges with residual delays that are above a specified threshold. By starting with a small residual threshold and proceeding to larger thresholds, we can generate a series of manifold visualizations that highlight various features of connectivity in the underlying network ({\em i.e.,} the US Internet).

\Cref{fig:public_us} shows the Internet latency space in the US generated by Matisse. The base map and the graph are sized and oriented so that locations of vertices in the graph are aligned appropriately on the map.  \Cref{fig:public_us} shows the manifold for connectivity at a low residual latency threshold of $10 \unit{\milli\second}$ ({\em i.e.,} links are only specified where residual delay is close to GCL).  This view shows areas of efficient connectivity along the East Coast and highlights the saddle regions associated with negative Ricci curvature assigned to links that span the Midwest and Southern regions. These regions exhibit locally rich intra-region fiber optic connections where neighboring nodes are typically directly linked to each other.

Matisse reveals the dual nature of connectivity: regions well-served by positive Ricci curvature connections illustrate a densely interconnected infrastructure, whereas negatively curved edges highlight geographic challenges and infrastructural limitations. These findings underscore the significant impact of geography on the network's performance in the US Internet.

\subsection{Manifolds Reflect the Delay Space}
\label{sec:manifold_distance}
To demonstrate that key network metrics are well-captured by our manifold view, we leverage the case of the US Internet described in~\cref{sec:internet_US} and show that our generated manifolds succeed in reflecting its hidden delay space.

\subsubsection{Accurate recovery of delays} We start by selecting pairs of vantage points (VPs) on our manifold and compute the geodesic distance as induced by the surface geometry of the manifold using \cite{Liu_2022}.
This enables us to identify both the distance between the cities hosting the VPs and a specific path between the VPs.
We then construct linear predictors for the latency as a function of either {\em (i)} the GCD or {\em (ii)} the geodesic distance. The best linear predictor is obtained by applying a linear regression on the different distances and acts as a scaling factor to express the geodesic distance in terms of latency.

Quantitatively, the relationship has a high coefficient of determination $r^2 \approx 0.7664$. This value is necessarily less than $1$ because triangle inequality violations that are quite common in Internet measurement datasets~\cite{Zheng2005InternetRP}. That is to say, a distance metric on a manifold cannot perfectly capture the Internet delay space. Removing those latencies that contribute to TIVs yields $r^2$ value $0.8827$.

\begin{table}[ht]
\centering
\caption{Geodesic prediction of Internet latency for city pairs in the US. $\varepsilon$ indicates the threshold where the edge appears for the first time; $\delta_\text{GCD}, \delta_\text{Geo}$ denote the difference between the estimated latency using the great circle distance and the geodesic distance and the observed latency, resp.; $d_{\text{GCD}}, d_{\text{RD}}$ denote the distance between points according to the great circle distance and long-haul cables as obtained in \cite{rama-map}. All times are measured in milliseconds.}
\label{tab:distance}
\small
\begin{tabular}{|c|c|c|c|c|c|c|c|}
\hline
$\varepsilon$ & City A     & City B     & $\delta_{\text{GCD}}$ & $\delta_{\text{Geo}}$ & $d_\text{RD}$ & $d_{\text{GCD}}$ \\ \hline
10            & Detroit    & Pittsburgh & 4.7                 & 1.4                   & 629.2         & 349.9            \\ \hline
10            & Ashburn    & Atlanta    & 9.6                 & 6.5                   & 1084.7        & 853.1            \\ \hline
16            & Phoenix    & Dallas     & -6.7                & -2.1                  & 1707.5        & 1418.2           \\ \hline
18            & St. George & Denver     & 10.8                & 2.4                   & 1021.9        & 808.5            \\ \hline
22            & Dallas     & L.A.       & 23.9                & 4.5                   & 2314.2        & 1993.6           \\ \hline
\end{tabular}
\end{table}

\begin{figure}[ht]
    \centering
    \includegraphics[width=0.8\columnwidth]{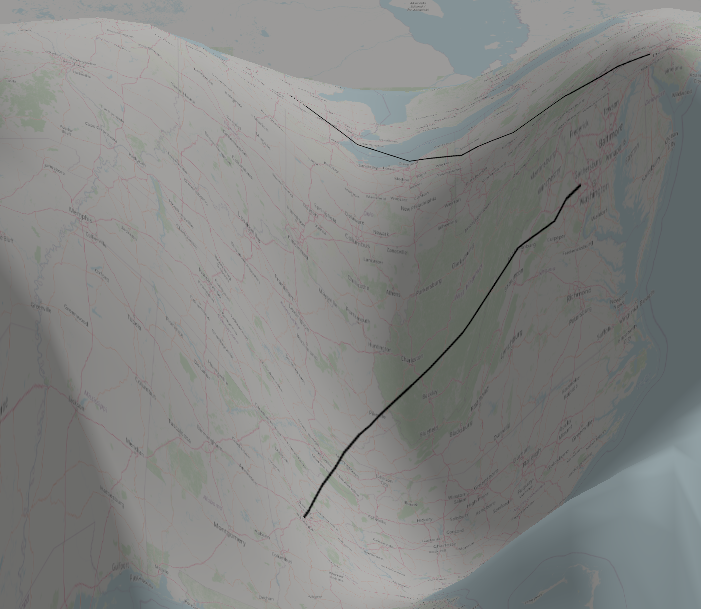}
    \caption{Zoom-in of Figure \cref{fig:public_us} showing the geodesic paths computed between Detroit and Pittsburgh and between Ashburn and Atlanta. The manifold uses $\varepsilon = 10 \unit{\milli\second}$. This overhead view shows the geodesics differ from line-of-sight paths.}
    \label{fig:public_us_geodesics}
    \vspace{-.25in}
\end{figure}

When we analyze the linear prediction of latency using the geodesic distance and the GCD, we find that both approaches yield similar $r^2$ values. However, a more detailed examination of the crucial connectivity bridges identified by sweeping through different thresholds reveals Matisse's ability to more faithfully model the latency of the paths crossing these vital links, as shown in \cref{tab:distance}. For the important links shown, we observe that $|\delta_{\text{Geo}}| < |\delta_{\text{GCD}}|$, meaning geodesic distance more accurately reflects the latencies for these links than GCD does.

Taking a broader perspective, this result suggests that at a certain threshold, the identification of saddle points that distort the space in a manner that more accurately reflects real-world Internet routing becomes critical for an accurate recovery of the distance taken. Consequently, paths traversing these saddles at the threshold at which they appear, as in \cref{fig:public_us_geodesics}, will on average enable more accurate estimates of latency compared to those calculated using GCD alone. This insight highlights the importance of sweeping through a range of thresholds of connectivity to derive latency estimates from manifolds that mirror the complex routing dynamics in the Internet.

\subsubsection{Stability} With regards to stability of the output, we find that the fitted geodesic distance between a fixed pair of vantage points does not vary much across a time span of weeks. This stability aligns with the minimal fluctuations we observed in the minimum RTT across that time-period. To test this, we gathered snapshots of data from the US every half day for two weeks. We then generated a manifold with Matisse and checked the fitted geodesic distances. In one particular case, the RTT estimated from the geodesic between Los Angeles and New York City stayed within $72.9$ and $75.3$ milliseconds, yielding a range of just $2.28$ milliseconds. Other city pairs featured similarly small ranges.

Clearly, the absence of a straight-line shortest path between two cities translates into a larger threshold of connectivity. Similarly, the sparsity of connections to bridge different strongly connected components of the networks results in negative curvature, as suggested by the toy example in \cref{fig:toy}. While our geodesics may not precisely describe the actual path taken by the data and reflected in the measurements, the tendency to pass through saddle regions highlights parts of the network that lack direct physical connectivity. That is, the geodesic distance is a more appropriate and intuitive estimator of latency than GCD because by having to move along the surface of the manifold rather than by using line-of-sight, it takes into account the dilatation of the delay space that occurs due to the absence of infrastructure. In conclusion, the geodesic properties demonstrate the ability of our generated manifolds to faithfully represent the Internet delay space as inferred from the original measurements.

\section{Summary and Future Work}  \label{sec:conclusion}    Visualizations of manifolds derived from real-world measurements can improve our understanding of the basic characteristics of those measurements and reveal important characteristics such as anomalies or other unusual aspects of the system where the measurements originated.  In this paper, we describe a methodology for generating manifold representations of planar graphs that include locations of vertices and Ricci curvature values on edges.  Our methodology generates a manifold with a smooth surface that inherits the curvature characteristics of the underlying graph and preserves the locations of vertices.  Importantly, the manifold enables recovery of the original measurements and inference of distances that were not included in the original data.  We realize our methodology in a system that we call {\em Matisse}, which generates manifold visualizations and can be applied iteratively and manipulated to assess systems over a range of topological configurations.  To demonstrate our method and Matisse, we present case studies.  These include a toy network that we use to illustrate the core aspects of Matisse's functionality.  To illustrate the utility and flexibility of the tool, we also apply Matisse to graphs derived from measured delays in the US public Internet.

In future work, we plan to use Matisse to examine other types of Internet measurements through their manifold representations. We also intend to explore using Matisse to improve our understanding of other geographically-embedded cyber-physical critical infrastructures such as transportation networks and energy networks.

\bibliographystyle{IEEEtran}
\bibliography{ref}

\begin{thebibliography}{10}
\providecommand{\url}[1]{#1}
\csname url@samestyle\endcsname
\providecommand{\newblock}{\relax}
\providecommand{\bibinfo}[2]{#2}
\providecommand{\BIBentrySTDinterwordspacing}{\spaceskip=0pt\relax}
\providecommand{\BIBentryALTinterwordstretchfactor}{4}
\providecommand{\BIBentryALTinterwordspacing}{\spaceskip=\fontdimen2\font plus
\BIBentryALTinterwordstretchfactor\fontdimen3\font minus \fontdimen4\font\relax}
\providecommand{\BIBforeignlanguage}[2]{{%
\expandafter\ifx\csname l@#1\endcsname\relax
\typeout{** WARNING: IEEEtran.bst: No hyphenation pattern has been}%
\typeout{** loaded for the language `#1'. Using the pattern for}%
\typeout{** the default language instead.}%
\else
\language=\csname l@#1\endcsname
\fi
#2}}
\providecommand{\BIBdecl}{\relax}
\BIBdecl

\bibitem{sigmetrics2022}
L.~Salamatian, S.~Anderson, J.~Matthews, P.~Barford, W.~Willinger, and M.~Crovella, ``{Curvature-based Analysis of Network Connectivity in Private Backbone Infrastructures},'' in \emph{{Proc. ACM Meas. Anal. Comput. Syst.}}, 2022.

\bibitem{cacm_23}
\BIBentryALTinterwordspacing
L.~Salamatian, S.~Anderson, J.~Mathews, P.~Barford, W.~Willinger, and M.~Crovella, ``A manifold view of connectivity in the private backbone networks of hyperscalers,'' \emph{Commun. ACM}, vol.~66, no.~8, p. 95–103, jul 2023. [Online]. Available: \url{https://doi.org/10.1145/3604620}
\BIBentrySTDinterwordspacing

\bibitem{wold1987principal}
S.~Wold, K.~Esbensen, and P.~Geladi, ``Principal component analysis,'' \emph{Chemometrics and intelligent laboratory systems}, vol.~2, no. 1-3, pp. 37--52, 1987.

\bibitem{vandermaaten08}
L.~van~der Maaten and G.~Hinton, ``{Visualizing Data Using t-SNE},'' \emph{{Journal of Machine Learning Research}}, vol.~9, p. 2579–2605, 2008.

\bibitem{mcinnis20}
\BIBentryALTinterwordspacing
L.~McInnes, J.~Healy, and J.~Melville, ``{UMAP: Uniform Manifold Approximation and Projection for Dimension Reduction},'' September 2020. [Online]. Available: \url{https://arxiv.org/abs/1802.03426}
\BIBentrySTDinterwordspacing

\bibitem{ollivier2009ricci}
Y.~Ollivier, ``Ricci curvature of markov chains on metric spaces,'' \emph{Journal of Functional Analysis}, vol. 256, no.~3, pp. 810--864, 2009.

\bibitem{Ni15}
C.-C. Ni, Y.-Y. Lin, J.~Gao, X.~D. Gu, and E.~Saucan, ``{Ricci Curvature of the Internet Topology},'' \emph{{arXiv}}, no. 1501.04138, 2015.

\bibitem{sun2010effectiveness}
H.~Sun and Z.~Li, ``Effectiveness of cartogram for the representation of spatial data,'' \emph{The Cartographic Journal}, vol.~47, no.~1, pp. 12--21, 2010.

\bibitem{cauvin1989cartographic}
C.~Cauvin and C.~Schneider, ``Cartographic transformations and the piezopleth maps method,'' \emph{The Cartographic Journal}, vol.~26, no.~2, pp. 96--104, 1989.

\bibitem{hoorn2023ollivier}
P.~v.~d. Hoorn, G.~Lippner, C.~Trugenberger, and D.~Krioukov, ``Ollivier curvature of random geometric graphs converges to ricci curvature of their riemannian manifolds,'' \emph{Discrete \& Computational Geometry}, vol.~70, no.~3, pp. 671--712, 2023.

\bibitem{Joshi2007EnergyMF}
P.~Joshi and C.~H. S{\'e}quin, ``Energy minimizers for curvature-based surface functionals,'' \emph{Computer-aided Design and Applications}, vol.~4, pp. 607--617, 2007.

\bibitem{2020SciPy-NMeth}
P.~Virtanen, R.~Gommers, T.~E. Oliphant, M.~Haberland, T.~Reddy, D.~Cournapeau, E.~Burovski, P.~Peterson, W.~Weckesser, J.~Bright, S.~J. {van der Walt}, M.~Brett, J.~Wilson, K.~J. Millman, N.~Mayorov, A.~R.~J. Nelson, E.~Jones, R.~Kern, E.~Larson, C.~J. Carey, {\.I}.~Polat, Y.~Feng, E.~W. Moore, J.~{VanderPlas}, D.~Laxalde, J.~Perktold, R.~Cimrman, I.~Henriksen, E.~A. Quintero, C.~R. Harris, A.~M. Archibald, A.~H. Ribeiro, F.~Pedregosa, P.~{van Mulbregt}, and {SciPy 1.0 Contributors}, ``{{SciPy} 1.0: Fundamental Algorithms for Scientific Computing in Python},'' \emph{Nature Methods}, vol.~17, pp. 261--272, 2020.

\bibitem{threejs}
{Ricardo Cabello (Mr.doob)}, ``{three.js},'' \url{https://github.com/mrdoob/three.js/}, First released in 2010, continuously updated, accessed: 2023-11.

\bibitem{openlayers}
``Openlayers api,'' \url{https://openlayers.org/}, 2021.

\bibitem{flamary2021pot}
\BIBentryALTinterwordspacing
R.~Flamary, N.~Courty, A.~Gramfort, M.~Z. Alaya, A.~Boisbunon, S.~Chambon, L.~Chapel, A.~Corenflos, K.~Fatras, N.~Fournier, L.~Gautheron, N.~T. Gayraud, H.~Janati, A.~Rakotomamonjy, I.~Redko, A.~Rolet, A.~Schutz, V.~Seguy, D.~J. Sutherland, R.~Tavenard, A.~Tong, and T.~Vayer, ``Pot: Python optimal transport,'' \emph{Journal of Machine Learning Research}, vol.~22, no.~78, pp. 1--8, 2021. [Online]. Available: \url{http://jmlr.org/papers/v22/20-451.html}
\BIBentrySTDinterwordspacing

\bibitem{matisseRepository}
S.~Jasina, J.~Mathews, and L.~Salamatian, \url{https://github.com/StephenJasina/linear-geodesic-optimization}, August 2025.

\bibitem{willinger2013}
W.~Willinger and M.~Roughan, ``Internet topology research redux,'' \emph{Recent Advances in Networking}, 2013.

\bibitem{huffaker2002}
B.~Huffaker, D.~Plummer, D.~Moore, and K.~Claffy, ``Topology discovery by active probing,'' \emph{Proceedings 2002 Symposium on Applications and the Internet (SAINT) Workshops}, 2002.

\bibitem{ping}
\BIBentryALTinterwordspacing
``ping (networking utility),'' 2021. [Online]. Available: \url{https://en.wikipedia.org/wiki/Ping_(networking_utility)}
\BIBentrySTDinterwordspacing

\bibitem{ripeatlas}
``Ripe atlas,'' \url{https://atlas.ripe.net/}, 2021.

\bibitem{rama-map}
R.~Durairajan, S.~Ghosh, X.~Tang, P.~Barford, and B.~Eriksson, ``{Internet Atlas: A Geographic Database of the Internet},'' in \emph{Proceedings of the 5th ACM Workshop on HotPlanet}.\hskip 1em plus 0.5em minus 0.4em\relax New York, NY, USA: Association for Computing Machinery, 2013.

\bibitem{Liu_2022}
Z.~Liu, \url{https://github.com/zishun/MeshUtility}, May 2022.

\bibitem{Zheng2005InternetRP}
\BIBentryALTinterwordspacing
H.~Zheng, E.~K. Lua, M.~R. Pias, and T.~G. Griffin, ``Internet routing policies and round-trip-times,'' in \emph{Passive and Active Network Measurement Conference}, 2005. [Online]. Available: \url{https://www.cl.cam.ac.uk/research/dtg/archived/files/publications/public/mp431/pam2005.pdf}
\BIBentrySTDinterwordspacing

\bibitem{mark2008computational}
M.~Berg, O.~Cheong, M.~Kreveld, and M.~Overmars, \emph{Computational Geometry: Algorithms and Applications}.\hskip 1em plus 0.5em minus 0.4em\relax Springer, 2008.

\bibitem{crane2013DGP}
K.~Crane, F.~de~Goes, M.~Desbrun, and P.~Schröder, ``Digital geometry processing with discrete exterior calculus,'' in \emph{ACM SIGGRAPH 2013 courses}, ser. SIGGRAPH '13.\hskip 1em plus 0.5em minus 0.4em\relax New York, NY, USA: ACM, 2013.

\bibitem{lin2011ricci}
Y.~Lin, L.~Lu, and S.-T. Yau, ``Ricci curvature of graphs,'' \emph{Tohoku Mathematical Journal, Second Series}, vol.~63, no.~4, pp. 605--627, 2011.

\end{thebibliography}

\newpage
\appendix
\subsection{Clustering}
\label{sec:clustering}
\begin{figure}[tbp]
    \centering
    \begin{subfigure}{\columnwidth}
        \centering
        \includegraphics[width=0.8\columnwidth]{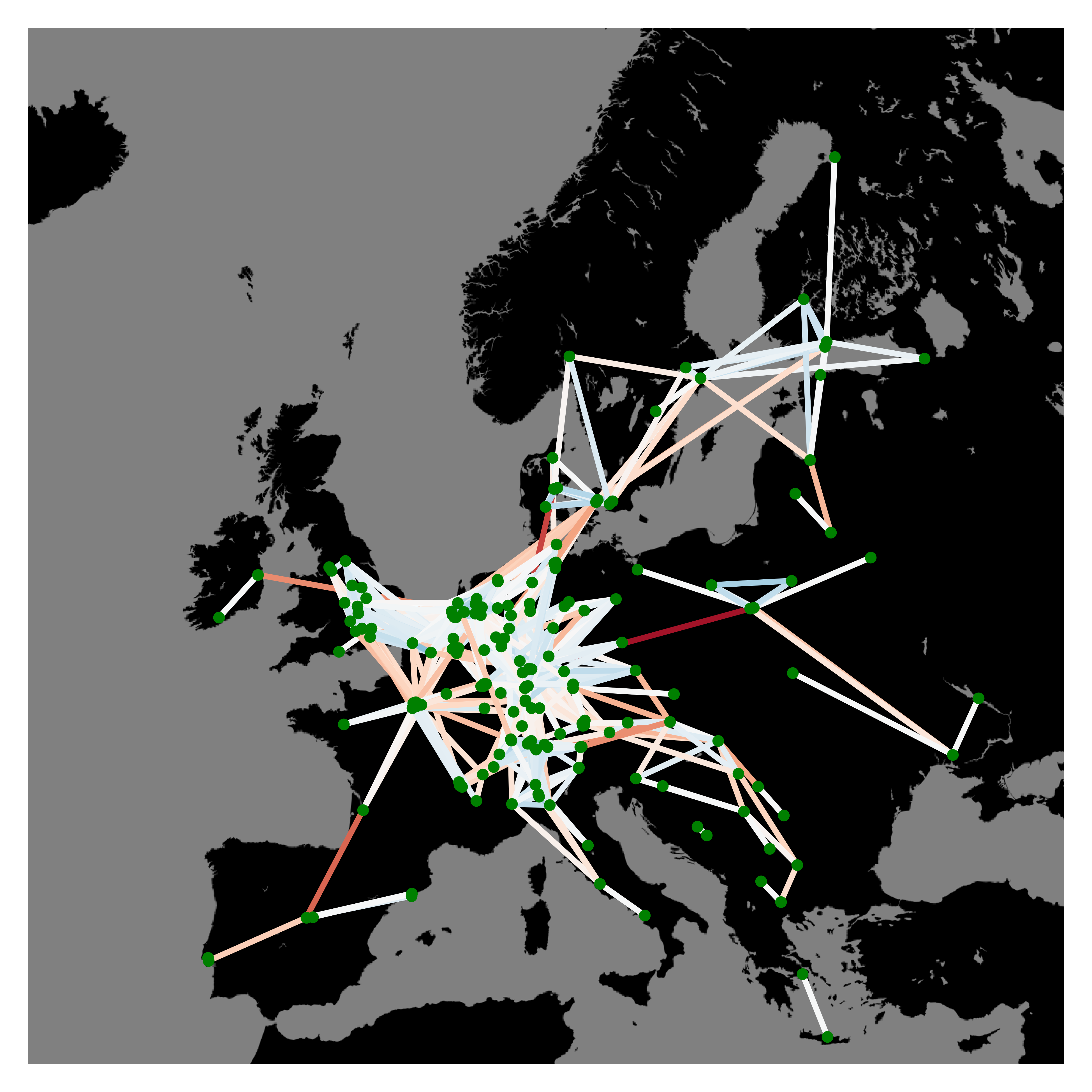}
        \label{fig:graph_Europe_unclustered}
    \end{subfigure}
    \begin{subfigure}{\columnwidth}
        \centering
        \includegraphics[width=\columnwidth]{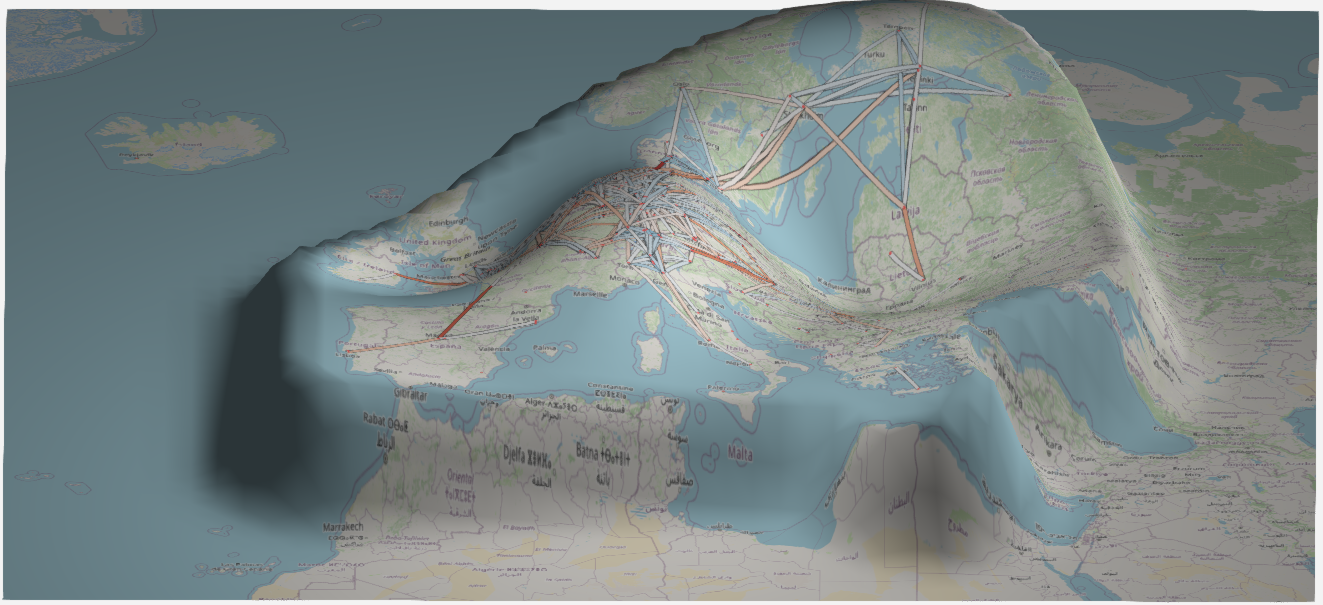}
        \label{fig:manifold_Europe_unclustered}
    \end{subfigure}
    \caption{On the top, the unclustered European dataset graph for a medium residual latency ($5$ ms), as well as the corresponding manifold on the bottom. For clarity, vertices with no incident edges are not displayed.}
    \label{fig:Europe_unclustered}
\end{figure}

\begin{figure}[tbp]
    \centering
    \begin{subfigure}{\columnwidth}
        \centering
        \includegraphics[width=0.8\columnwidth]{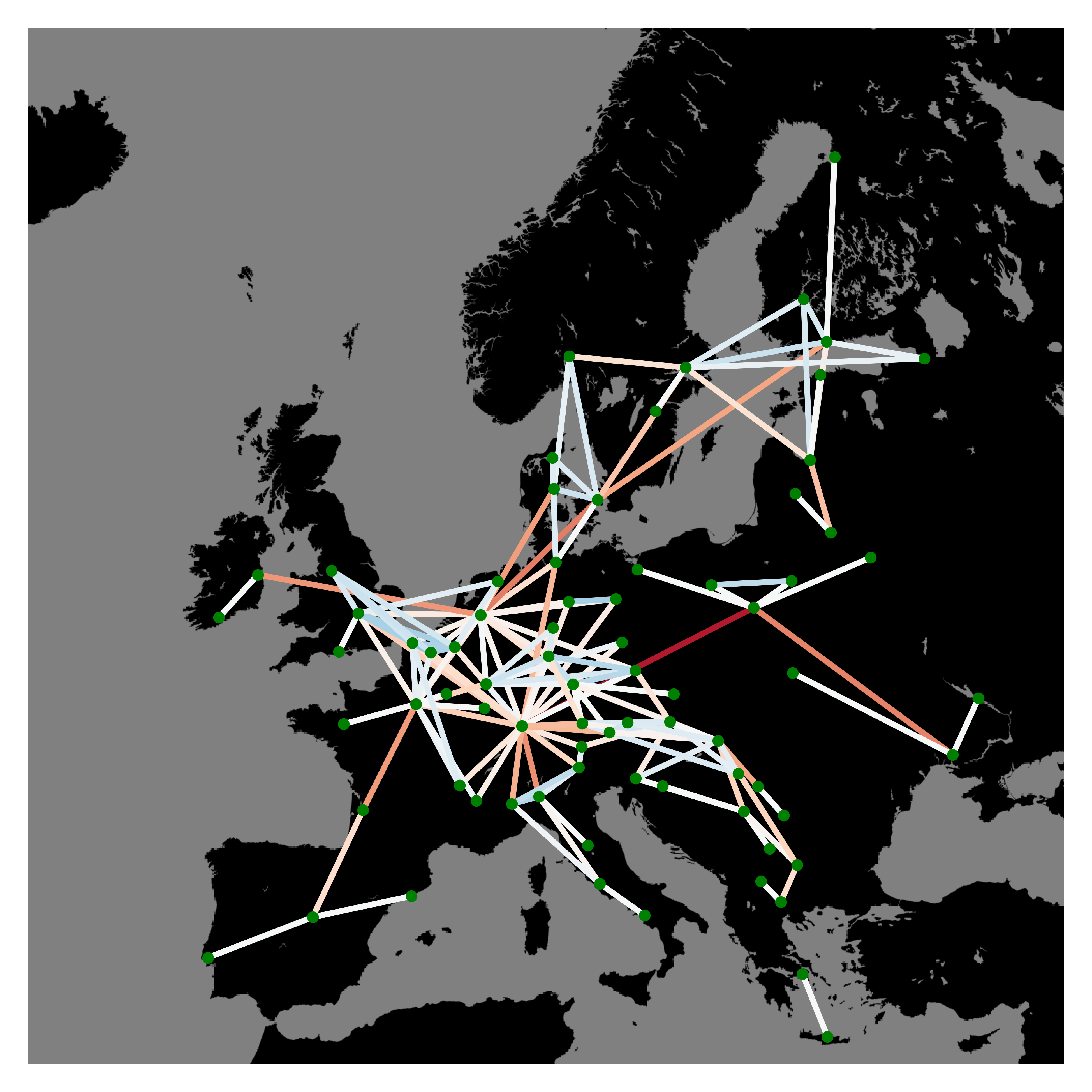}
        \caption{Europe public Internet after clustering with a distance threshold of $500$ kilometers.}
        \label{fig:graph_Europe_clustered_500}
    \end{subfigure}
    \begin{subfigure}{\columnwidth}
        \centering
        \includegraphics[width=0.8\columnwidth]{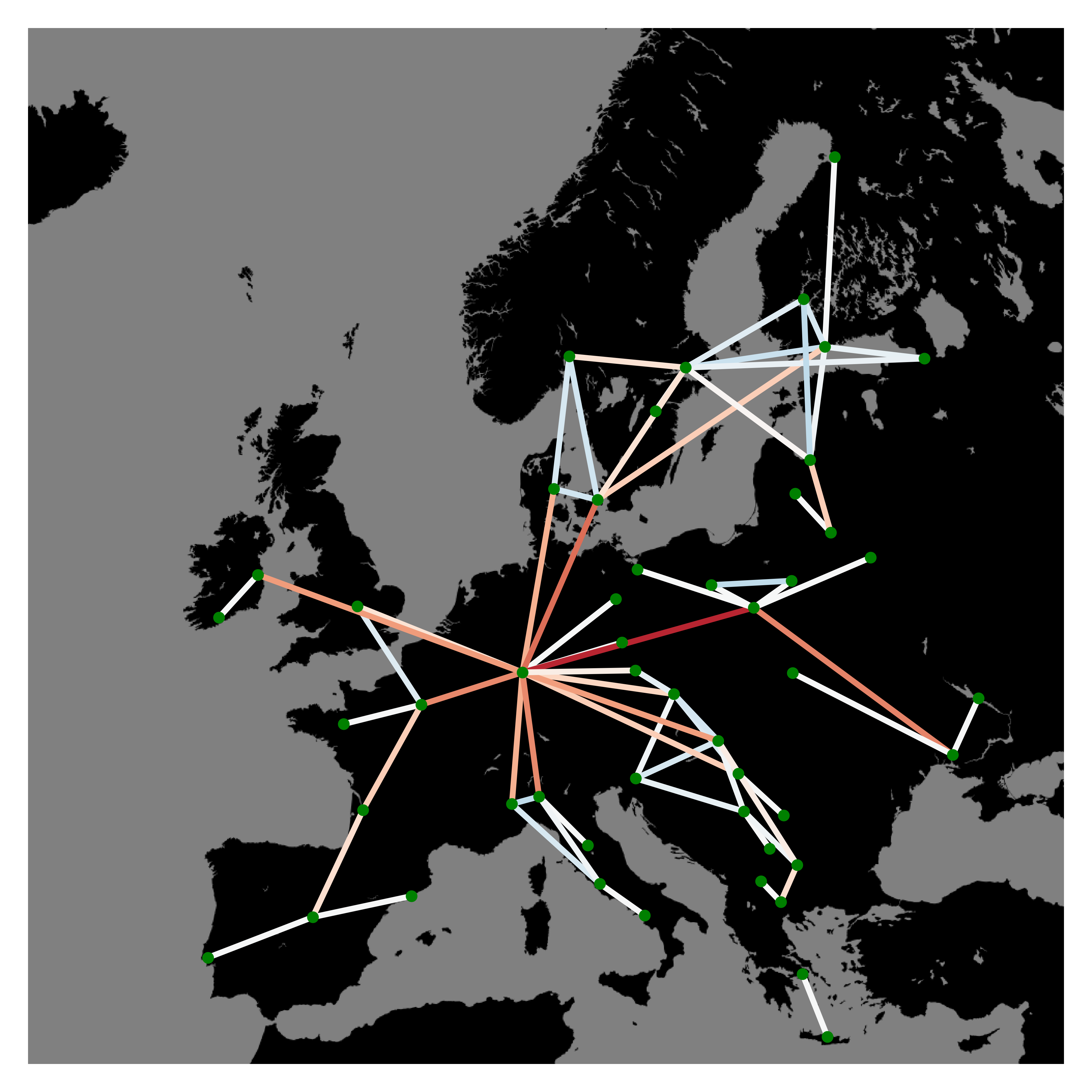}
        \caption{Europe public Internet after clustering with a distance threshold of $750$ kilometers.}
        \label{fig:graph_Europe_clustered_750}
    \end{subfigure}
    \caption{Examples of clustering run on the RIPE Atlas European public Internet dataset with $5$ millisecond residual delay threshold. For clarity, vertices with no incident edges are not displayed.}
    \label{fig:graph_Europe_clustered}
\end{figure}

In the main body of the paper, we highlighted the fact that very dense networks may necessitate preliminary clustering for practical analysis and effective visualization.  For example, without this step, using RIPE Atlas anchors in Europe results in the network and manifold depicted in \cref{fig:Europe_unclustered}. This representation reveals a considerable concentration of edges in the vicinity of France, Germany, and Switzerland, which results in a graph and manifold that are difficult to decipher. In particular, the optimization process leads to the formation of a hill with positive curvature that tends to dominate the visualization, despite the presence of many links characterized by negative curvature.

\Cref{fig:graph_Europe_clustered} shows the results of clustering the European Internet data at two different distance thresholds. Importantly, macroscopic features of the network such as long links that connect dense areas of connectivity are preserved by the clustering, and the general clutter is significantly reduced. Of particular note is the hub-like connectivity near Luxembourg and the tree-like structure near Ukraine.  Both of these features have important implications for network operations in terms of robustness and reachability.

\subsection{Technical Details of Computation}
\label{sec:details_of_computation}
This section details the strategy for computing important quantities on a mesh, like curvature, as well as the loss functions based on these quantities. Furthermore, details of computing gradients of these quantities are presented.

Due to the necessity of much more notation than in the main body of this paper, some choices of letters will differ. The exact notation used is listed in \cref{tab:variables}.

\begin{table*}[t]
    \centering
    \caption{Variables used for computing the objective function. Additionally listed are the runtime complexities for computing the variables across the entire mesh (\emph{e.g.}, across all choices of \(i\) and \(j\)). Quantities that are only computed once in the optimization have their runtimes omitted.}
    \label{tab:variables}
    \begin{tabular}{r|l|l}
        Variable & Definition & Runtime Complexity \\ \hline
        \(N_{i, j}\) & Outward normal of triangle \(v_i \to v_j \to v_{\nxt\pof{i, j}}\) & \(O\pof{\abs{V_M}}\) \\
        \(A_{i, j}\) & Area of triangle \(v_i \to v_j \to v_{\nxt\pof{i, j}}\) & \(O\pof{\abs{V_M}}\) \\
        \(D_{i, j}\) & Vertex triangle areas; diagonal & \(O\pof{\abs{V_M}}\) \\
        \(\theta_{i, j}\) & Measure of \(\angle v_iv_{\nxt\pof{i, j}}v_j\) & \(O\pof{\abs{V_M}}\) \\
        \(L_C^{\text{N}}\) & Cotangent operator with zero-Neumann boundary condition & \(O\pof{\abs{V_M}}\) \\
        \(L_C^{\text{D}}\) & Cotangent operator with zero-Dirichlet boundary condition & \(O\pof{\abs{V_M}}\) \\
        && \\
        \(E_G^\epsilon\) & The set of network edges at threshold \(\epsilon\) & \\
        \(\kappa^\text{R}_e\) & The Ollivier-Ricci curvature of the edge \(E_G^\epsilon\) & \\
        \(\widetilde{\kappa^\text{G}}_i\) & The discrete Gaussian curvature at \(v_i\), scaled by vertex area & \(O\pof{\abs{V_M}}\) \\
        \(\kappa^\text{G}_i\) & The discrete Gaussian curvature at \(v_i\) & \(O\pof{\abs{V_M}}\) \\
        \(\widetilde{N}_i\) & An outward pointing vector at \(v_i\) & \(O\pof{\abs{V_M}}\) \\
        \(\widetilde{\kappa^\text{H}}_i\) & The mean curvature normal at \(v_i\) & \(O\pof{\abs{V_M}}\) \\
        \(\kappa^\text{H}_i\) & The mean curvature at \(v_i\) & \(O\pof{\abs{V_M}}\) \\
        \(\kappa^+_i\) & The first principal curvature at \(v_i\) & \(O\pof{\abs{V_M}}\) \\
        \(\kappa^-_i\) & The second principal curvature at \(v_i\) & \(O\pof{\abs{V_M}}\) \\
        && \\
        \(B_r\pof{e}\) & The ball of radius \(r\) around \(e\) & \\
        \(\mathcal{L}_{\mathrm{curvature}}\pof{M}\) & Sum of squares of the differences between vertices actual and desired curvatures & \makecell[lt]{\(O\pof{\abs{E_G} \cdot \abs{V_M}}\) \\ (often \(O\pof{\abs{E_G} \cdot \sqrt{\abs{V_M}}}\))} \\
        && \\
        \(\mathcal{L}_{\mathrm{smooth}}\pof{M}\) & The surface area independent \(\text{MVS}_{\text{cross}}\) energy~\cite{Joshi2007EnergyMF} & \(O\pof{\abs{V_M}}\)
    \end{tabular}
\end{table*}

\subsubsection{Inputs and Outputs}
As input to the optimization process, we take an weighted undirected graph \(G = \pof{V_G, E_G}\). Each vertex \(s \in V_G\) represents a node in a network and is annotated with location information. We therefore can compute quantities like \(\GCL\pof{s, s'}\), the Great Circle Latency between \(s\) and \(s'\).

Additionally, an edge \(\cof{s, s'} \in E_G\) has an associated measured Round Trip Time \(\RTT\pof{s, s'}\). In practice, latencies are collected for almost every pair of nodes, so \(G\) is nearly complete.

We also take several hyperparameters. First is the residual latency threshold \(\epsilon\). Next is \(\lambda_{\text{smooth}}\), which is the weighting parameter for the smoothness component of the loss function. Finally, there are a few hyperparameters describing the structure of the mesh, such as the number of triangles.

We return a triangle mesh \(M = \pof{V_M, E_M}\), stored in a doubly connected edge list format. That is, each edge is actually stored as a pair of directed edges, except for on the boundary, where only a single directed edge is used. These directed edges trace out each face of the mesh counterclockwise. An introduction to this data structure can be found in~\cite{mark2008computational}.

As hinted at above, the actual vertex-edge connectivity of the mesh is to be selected before running the optimization. That said, each vertex has coordinates in \(\mathbb{R}^3\) which are to be chosen by the optimization algorithm. For the purposes of the algorithm and mesh regularity, we parameterize each vertex position with a single number. In our current implementation, each vertex \(v \in V_M\) can be broken into parts as \(\pof{p_v, z_v}\), where \(p_v\) is a latitude-longitude pair, and \(z_v\) is an altitude. The optimization algorithm then determines the best \(z\) values.

In the upcoming sections, the computations can take serious advantage of vectors and matrices. Therefore, while notationally inelegant, we will assign indices to the vertices in \(V_M\).

On that note, if \(i\) and \(j\) are two indices for which \(\pof{v_i, v_j} \in E_M\), let \(\nxt\pof{i, j}\) be the index such that \(v_i \to v_j \to v_{\nxt\pof{i, j}}\) traces a triangle counterclockwise. If \(\nxt\pof{i, j}\) does not exist, then the half-edge \(\pof{v_i, v_j}\) lies on the boundary.

We also write \(\partial M\) to represent the boundary of our mesh. Abusing notation, we can write \(v_i \in \partial M\) and \(\pof{v_i, v_j} \in \partial M\) to denote that a vertex or an edge is on the boundary, respectively.

\subsubsection{Laplacian}
We use the standard definition to compute the cotangent Laplacian on a mesh~\cite{crane2013DGP}.

\paragraph{Forward Computation}
First, we compute the following useful quantities:
\begin{align*}
    N_{i, j} &= \pof{v_i - v_{\nxt\pof{i, j}}} \times \pof{v_j - v_{\nxt\pof{i, j}}}, \\
    A_{i, j} &= \frac{1}{2}\norm{N_{i, j}}_2, \\
    D_{i, j} &= \begin{cases}
        \frac{1}{3}{\sum_{\substack{k \\ \pof{v_i, v_k} \in E_M}}A_{i, k}} & \text{if \(i = j\)}, \\
        0 & \text{otherwise},
    \end{cases} \\
    \cot\pof{\theta_{i, j}} &= \frac{\pof{v_i - v_{\nxt\pof{i, j}}} \cdot \pof{v_j - v_{\nxt\pof{i, j}}}}{2A_{i, j}}.
\end{align*}

Now, we compute the Laplacian matrices. We must break into cases: \begin{itemize}
    \item
    If \(\pof{v_i, v_j} \in \partial M\), then \(\pof{L_C^{\text{N}}}_{i, j} = \frac{1}{2}\cot\pof{\theta_{i, j}}\).
    \item
    If \(\pof{v_j, v_i} \in \partial M\), then \(\pof{L_C^{\text{N}}}_{i, j} = \frac{1}{2}\cot\pof{\theta_{j, i}}\).
    \item
    If \(\pof{v_i, v_j} \in E_M\) and \(\pof{v_j, v_i} \in E_M\), then \[\pof{L_C^{\text{N}}}_{i, j} = \frac{1}{2}\pof{\cot\pof{\theta_{i, j}} + \cot\pof{\theta_{j, i}}}.\]
    \item
    If \(i = j\), \begin{alignat*}{1}
        \pof{L_C^{\text{N}}}_{i, j} &= \frac{1}{2}\sum_{\substack{k \\ \pof{v_i, v_k} \in E_M}}\cot\pof{\theta_{i, k}} \\
        &\qquad+ \frac{1}{2}\sum_{\substack{k \\ \pof{v_k, v_i} \in E_M}}\cot\pof{\theta_{k, i}}.
    \end{alignat*}
    \item
    Otherwise, \(\pof{L_C^{\text{N}}}_{i, j} = 0\).
\end{itemize}

Similarly, \begin{itemize}
    \item
    If \(\pof{v_i, v_j} \in E_M\), \(v_i \not\in \partial M\), and \(v_j \not\in \partial M\), then \[\pof{L_C^{\text{D}}}_{i, j} = \frac{1}{2}\pof{\cot\pof{\theta_{i, j}} + \cot\pof{\theta_{j, i}}}.\]
    \item
    If \(i = j\) and \(v_i \not\in \partial M\), then \begin{alignat*}{1}
        \pof{L_C^{\text{D}}}_{i, j} &= -\frac{1}{2}{\sum_{\substack{k \not\in \partial M \\ \pof{v_i, v_k} \in E_M \\ \pof{v_k, v_i} \in E_M}}\pof{\cot\pof{\theta_{i, k}} + \cot\pof{\theta_{k, i}}}}
    \end{alignat*}
    \item
    Otherwise, \(\pof{L_C^{\text{D}}}_{i, j} = 0.\)
\end{itemize}

\paragraph{Reverse Computation}
We now compute the partial derivatives of the above quantities. Here and in future sections, the key technique is to just apply the chain rule.

First, we have \begin{align*}
    \frac{\partial v_i}{\partial z_\ell} &= \begin{cases}
        e_3 & \text{if \(\ell = i\)}, \\
        0 & \text{otherwise},
    \end{cases} \\
    \frac{\partial N_{i, j}}{\partial z_\ell} &= \begin{cases}
        \pof{v_{\nxt\pof{i, j}} - v_j} \times \frac{\partial v_\ell}{\partial z_\ell} & \text{if \(\ell = i\)}, \\
        \pof{v_i - v_{\nxt\pof{i, j}}} \times \frac{\partial v_\ell}{\partial z_\ell} & \text{if \(\ell = j\)}, \\
        \pof{v_j - v_i} \times \frac{\partial v_\ell}{\partial z_\ell} & \text{if \(\ell = \nxt\pof{i, j}\)}, \\
        0 & \text{otherwise},
    \end{cases} \\
    \frac{\partial A_{i, j}}{\partial z_\ell} &= \frac{1}{4A_{i, j}}N_{i, j} \cdot \frac{\partial N_{i, j}}{\partial z_\ell}, \\
    \pof{\frac{\partial D}{\partial z_\ell}}_{i, j} &= \begin{cases}
        \frac{1}{3}{\sum_{\substack{k \\ \pof{v_i, v_k} \in E_M}}\frac{\partial A_{i, k}}{\partial z_\ell}} & \text{if \(i = j\)}, \\
        0 & \text{otherwise},
    \end{cases}
\end{align*}

For cotangents, we have four cases. \begin{itemize}
    \item
    If \(\ell = i\), then \[\frac{\partial}{\partial z_\ell}\cot\pof{\theta_{i, j}} = \frac{\pof{v_j - v_{\nxt\pof{i, j}}} \cdot \frac{\partial v_\ell}{\partial z_\ell} - 2\cot\pof{\theta_{i, j}}\frac{\partial A_{i, j}}{\partial z_\ell}}{2A_{i, j}}.\]
    \item
    If \(\ell = j\), then \[\frac{\partial}{\partial z_\ell}\cot\pof{\theta_{i, j}} = \frac{\pof{v_i - v_{\nxt\pof{i, j}}} \cdot \frac{\partial v_\ell}{\partial z_\ell} - 2\cot\pof{\theta_{i, j}}\frac{\partial A_{i, j}}{\partial z_\ell}}{2A_{i, j}}.\]
    \item
    If \(\ell = \nxt\pof{i, j}\), then \[\frac{\partial}{\partial z_\ell}\cot\pof{\theta_{i, j}} = \frac{\pof{2v_{\nxt\pof{i, j}} - v_i - v_j} \cdot \frac{\partial v_\ell}{\partial z_\ell} - 2\cot\pof{\theta_{i, j}}\frac{\partial A_{i, j}}{\partial z_\ell}}{2A_{i, j}}.\]
    \item
    Otherwise, \(\frac{\partial}{\partial z_\ell}\cot\pof{\theta_{i, j}} = 0\).
\end{itemize}

The partials of the Laplacian matrix entries similarly have cases: \begin{itemize}
    \item
    If \(\pof{v_i, v_j} \in \partial M\), then \[\pof{\frac{\partial L_C^{\text{N}}}{\partial z_\ell}}_{i, j} = \frac{1}{2}\frac{\partial}{\partial z_\ell}\cot\pof{\theta_{i, j}}.\]
    \item
    If \(\pof{v_j, v_i} \in \partial M\), then \[\pof{\frac{\partial L_C^{\text{N}}}{\partial z_\ell}}_{i, j} = \frac{1}{2}\frac{\partial}{\partial z_\ell}\cot\pof{\theta_{j, i}}.\]
    \item
    If \(\pof{v_i, v_j} \in E_M\) and \(\pof{v_j, v_i} \in E_M\), then \[\pof{\frac{\partial L_C^{\text{N}}}{\partial z_\ell}}_{i, j} = \frac{1}{2}\pof{\frac{\partial}{\partial z_\ell}\cot\pof{\theta_{i, j}} + \frac{\partial}{\partial z_\ell}\cot\pof{\theta_{j, i}}}.\]
    \item
    If \(i = j\), then \begin{alignat*}{1}
        \pof{\frac{\partial L_C^{\text{N}}}{\partial z_\ell}}_{i, j} &= -\frac{1}{2}\sum_{\substack{k \\ \pof{v_i, v_k} \in E_M}}\frac{\partial}{\partial z_\ell}\cot\pof{\theta_{i, k}} \\
        &\qquad- \frac{1}{2}\sum_{\substack{k \\ \pof{v_k, v_i} \in E_M}}\frac{\partial}{\partial z_\ell}\cot\pof{\theta_{k, i}}.
    \end{alignat*}
    \item
    Otherwise, \(\pof{\frac{\partial L_C^{\text{N}}}{\partial z_\ell}}_{i, j} = 0\).
\end{itemize} Additionally: \begin{itemize}
    \item
    If \(\pof{v_i, v_j} \in E_M\), \(v_i \not\in \partial M\), and \(v_j \not\in \partial M\), then \[\pof{\frac{\partial L_C^{\text{D}}}{\partial z_\ell}}_{i, j} = \frac{1}{2}\pof{\frac{\partial}{\partial z_\ell}\cot\pof{\theta_{i, j}} + \frac{\partial}{\partial z_\ell}\cot\pof{\theta_{j, i}}}.\]
    \item
    If \(i = j\) and \(v_i \not\in \partial M\), then \[\pof{\frac{\partial L_C^{\text{D}}}{\partial z_\ell}}_{i, j} = -\frac{1}{2}{\sum_{\substack{k \not\in \partial M \\ \pof{v_i, v_k} \in E_M \\ \pof{v_k, v_i} \in E_M}}\pof{\frac{\partial}{\partial z_\ell}\cot\pof{\theta_{i, k}} + \frac{\partial}{\partial z_\ell}\cot\pof{\theta_{k, i}}}}.\]
    \item
    Otherwise, \(\pof{\frac{\partial L_C^{\text{D}}}{\partial z_\ell}}_{i, j} = 0\).
\end{itemize}

\subsubsection{Curvature}
For these computations (particularly the mean curvature one), consider \(v\) as a matrix of vertex positions, where each row corresponds to a vertex (so \(v\) is of shape \(\abs{V_M} \times 3\)). We will also use \(e_i \in \mathbb{R}^3\) to denote the \(i\)th standard basis vector.

\paragraph{Ollivier-Ricci Curvature}
Following the ideas from Lin, Lu, and Yau~\cite{lin2011ricci}, we use the POT library~\cite{flamary2021pot} to compute the Ollivier-Ricci curvatures of the edges of the graph \(\pof{V_G, E_G^\epsilon}\). Here, \(E_G^\epsilon \subseteq E_G\) is the set of edges whose RTTs are at most \(\epsilon\) milliseconds higher than their GCLs.

Note that \(\kappa^\text{R}_e\) is then only defined for edges in \(E_G^\epsilon\), as opposed to being defined for all edges in \(E_G\). We elide the \(\epsilon\) to reduce notational density.

\paragraph{Forward Computation}
We have \begin{align*}
    \theta_{i, j} &= \arctan\pof{\frac{1}{\cot\pof{\theta_{i, j}}}} \bmod \pi, \\
    \widetilde{\kappa^\text{G}}_i &= 2\pi - \sum_{\substack{k \\ \pof{v_i, v_k} \in E_M}} \theta_{k, c\pof{i, k}}, \\
    \kappa^\text{G} &= D^{-1}\widetilde{\kappa^\text{G}}, \\
    \widetilde{N}_i &= \sum_{\substack{k \\ \pof{v_i, v_k} \in E_M}} N_{i, k}, \\
    \widetilde{\kappa^\text{H}}_i &= -\frac{1}{2}e_i^\intercal D^{-1}L_C^{\text{N}}v, \\
    \kappa^\text{H}_i &= \sgn\pof{\widetilde{N}_i^\intercal\widetilde{\kappa^\text{H}}_i}\norm{\widetilde{\kappa^\text{H}}_i}_2, \\
    \kappa^+_i &= \kappa^{\text{H}}_i + \sqrt{\pof{\kappa^{\text{H}}_i}^2 - \kappa^{\text{G}}_i}, \\
    \kappa^-_i &= \kappa^{\text{H}}_i - \sqrt{\pof{\kappa^{\text{H}}_i}^2 - \kappa^{\text{G}}_i}.
\end{align*} The formulas for Gaussian curvature and mean curvature are standard from computational geometry~\cite{crane2013DGP}.

\paragraph{Reverse Computation}
Differentiating, \begin{align*}
    \frac{\partial\theta_{i, j}}{\partial z_\ell} &= -\frac{\partial\cot\pof{\theta_{i, j}}}{\partial z_\ell} \cdot \frac{1}{1 + \cot^2\pof{\theta_{i, j}}}, \\
    \frac{\partial\widetilde{\kappa^\text{G}}_i}{\partial z_\ell} &= -\sum_{\substack{k \\ \pof{v_i, v_k} \in E_M}} \frac{\partial\theta_{k, c\pof{i, k}}}{\partial z_\ell}, \\
    \frac{\partial\kappa^\text{G}}{\partial z_\ell} &= D^{-1}\pof{\frac{\dif\widetilde{\kappa^\text{G}}}{\partial z_\ell} - \frac{\dif D}{\partial z_\ell}\kappa^\text{G}}, \\
    \frac{\partial\widetilde{N}_i}{\partial z_\ell} &= \sum_{\substack{k \\ \pof{v_i, v_k} \in E_M}} \frac{\partial N_{i, k}}{\partial z_\ell}, \\
    \frac{\partial\widetilde{\kappa^\text{H}}_i}{\partial z_\ell} &= -\frac{1}{2}e_i^\intercal D^{-1}\pof{\pof{\frac{\partial L_C^{\text{N}}}{\partial z_\ell} - \frac{\partial D}{\partial z_\ell}D^{-1}L_C^{\text{N}}}v + L_C^{\text{N}}\frac{\partial v}{\partial z_\ell}}, \\
    \frac{\partial\kappa_i^\text{H}}{\partial z_\ell} &= \frac{\sgn\pof{\widetilde{N}_i^\intercal\widetilde{\kappa^\text{H}}_i}}{\norm{\widetilde{\kappa^\text{H}}_i}_2}\widetilde{\kappa^\text{H}}_i^\intercal\frac{\partial\widetilde{\kappa^\text{H}}_i}{\partial z_\ell}, \\
    \frac{\partial\kappa^+_i}{\partial z_\ell} &= \frac{2\kappa^+_i\frac{\partial\kappa^{\text{H}}_i}{\partial\rho_i} - \frac{\partial\kappa^{\text{G}}_i}{\partial\rho_i}}{\kappa^+_i - \kappa^-_i}, \\
    \frac{\partial\kappa^-_i}{\partial z_\ell} &= \frac{\frac{\partial\kappa^{\text{G}}_i}{\partial\rho_i} - 2\kappa^-_i\frac{\partial\kappa^{\text{H}}_i}{\partial\rho_i}}{\kappa^+_i - \kappa^-_i}.
\end{align*}

\subsubsection{Curvature Loss}
Before tackling the loss functional, we must determine what it means for a point to be close to an edge.

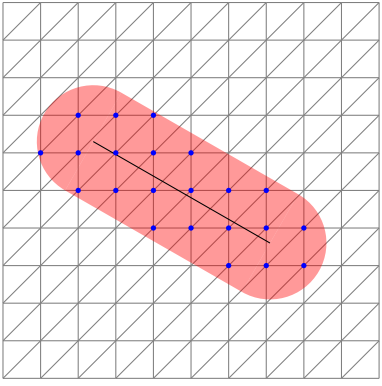
\begin{figure}[ht]
    \centering
    \begin{tikzpicture}[scale=0.5]
        % Draw the mesh
        \foreach \i in {0,...,9}{
            \foreach \j in {0,...,9}{
                \draw[color=gray] (\i,\j)--(\i+1,\j);
                \draw[color=gray] (\i,\j)--(\i,\j+1);
                \draw[color=gray] (\i,\j)--(\i+1,\j+1);
            }
            \draw[color=gray] (\i,10)--(\i+1,10) ;
            \draw[color=gray] (10,\i)--(10,\i+1);
        }

        % Draw the outline of the ball around the edge
        \fill[color=red, opacity=0.4] ({2.4+1.5*cos(atan((7.1-2.4)/(6.3-3.6)))},{6.3+1.5*sin(atan((7.1-2.4)/(6.3-3.6)))}) arc ({atan((7.1-2.4)/(6.3-3.6))}:{atan((7.1-2.4)/(6.3-3.6))+180}:1.5);
        \fill[color=red, opacity=0.4] ({7.1-1.5*cos(atan((7.1-2.4)/(6.3-3.6)))},{3.6-1.5*sin(atan((7.1-2.4)/(6.3-3.6)))}) arc ({atan((7.1-2.4)/(6.3-3.6))-180}:{atan((7.1-2.4)/(6.3-3.6))}:1.5);
        \fill[color=red, opacity=0.4] ({2.4-(1.5*(6.3-3.6)/sqrt((2.4-7.1)^2+(6.3-3.6)^2))},{6.3+(1.5*(2.4-7.1)/sqrt((2.4-7.1)^2+(6.3-3.6)^2))}) -- ({7.1-(1.5*(6.3-3.6)/sqrt((2.4-7.1)^2+(6.3-3.6)^2))},{3.6+(1.5*(2.4-7.1)/sqrt((2.4-7.1)^2+(6.3-3.6)^2))}) -- ({7.1+(1.5*(6.3-3.6)/sqrt((2.4-7.1)^2+(6.3-3.6)^2))},{3.6-(1.5*(2.4-7.1)/sqrt((2.4-7.1)^2+(6.3-3.6)^2))}) -- ({2.4+(1.5*(6.3-3.6)/sqrt((2.4-7.1)^2+(6.3-3.6)^2))},{6.3-(1.5*(2.4-7.1)/sqrt((2.4-7.1)^2+(6.3-3.6)^2))});

        % Draw the edge
        \draw (2.4,6.3)--(7.1,3.6);

        % Draw the points in the ball. Do this manually because automatic is too hard
        \fill[color=blue] (1,6) circle (2pt);
        \fill[color=blue] (2,5) circle (2pt);
        \fill[color=blue] (2,6) circle (2pt);
        \fill[color=blue] (2,7) circle (2pt);
        \fill[color=blue] (3,5) circle (2pt);
        \fill[color=blue] (3,6) circle (2pt);
        \fill[color=blue] (3,7) circle (2pt);
        \fill[color=blue] (4,4) circle (2pt);
        \fill[color=blue] (4,5) circle (2pt);
        \fill[color=blue] (4,6) circle (2pt);
        \fill[color=blue] (4,7) circle (2pt);
        \fill[color=blue] (5,4) circle (2pt);
        \fill[color=blue] (5,5) circle (2pt);
        \fill[color=blue] (5,6) circle (2pt);
        \fill[color=blue] (6,3) circle (2pt);
        \fill[color=blue] (6,4) circle (2pt);
        \fill[color=blue] (6,5) circle (2pt);
        \fill[color=blue] (7,3) circle (2pt);
        \fill[color=blue] (7,4) circle (2pt);
        \fill[color=blue] (7,5) circle (2pt);
        \fill[color=blue] (8,3) circle (2pt);
        \fill[color=blue] (8,4) circle (2pt);
    \end{tikzpicture}
    \caption{An overhead visualization of $B_r\pof{e}$. The mesh is in gray, $\pi\pof{e}$ is in black, and $B_r\pof{e}$ is in blue.}
    \label{fig:cuvature_loss_edge_ball}
\end{figure}

\paragraph{Determining the Ball Around an Edge}
Suppose \(s\) and \(s'\) are vertices in \(V_G\), and let \(e = \cof{s, s'}\). Recall that \(p_s\) (similarly \(p_{s'}\)) is the point in \(\mathbb{R}^2\) corresponding to \(s\) (similarly \(s'\)). Define \(\pi\pof{e}\) to be the edge connecting \(p_s\) and \(p_{s'}\). We say that \(v \in B_r\pof{e}\) when \(p_v\) is within distance \(r\) of \(\pi\pof{e}\). \Cref{fig:cuvature_loss_edge_ball} shows this setup.

To determine whether a vertex is in the ball, we break the problem into three parts: \begin{itemize}
    \item
    Determine whether \(p_v \in B_r\pof{p_s}\). This is the case when \(\norm{p_v - p_s}_2 < r\).
    \item
    Determine whether the distance from \(p_v\) to \(\pi\pof{e}\) is less than \(r\). To do this, we first project \(p_v\) onto the line containing \(\pi\pof{e}\) to get \(p_{v, e}\): \[p_{v, e} = p_s + \frac{\pof{p_{s'} - p_s}^\intercal\pof{p_v - p_s}}{\norm{p_{s'} - p_s}_2^2}\pof{p_{s'} - p_s}.\] To determine whether this projection lies on the actual segment and that the projection is not too far away from the original point, we simultaneously check the three conditions \begin{align*}
        \pof{p_v - p_s}^\intercal\pof{p_{s'} - p_s} \ge 0, \\
        \pof{p_v - p_{s'}}^\intercal\pof{p_s - p_{s'}} \ge 0, \\
        \norm{p_v - p_{v, e}}_2 < r.
    \end{align*}
    \item
    Determine whether \(p_v \in B_r\pof{p_{s'}}\). This is the case when \(\norm{p_v - p_{s'}}_2 < r\).
\end{itemize} If any of the three above parts yields a positive response, then \(v \in B_r\pof{e}\).

Note that the number of points in \(B_r\pof{p_{s'}}\) is typically \(O\pof{\sqrt{\abs{V_M}}}\), as long as \(r\) is chosen to be sufficiently small. In our work, we select \(r\) so that the tube around the edge is as thin as possible without disconnecting the tube (see \cref{sec:parameters}).

\paragraph{Forward Computation}
We define \[\mathcal{L}_{\mathrm{curvature}}\pof{M} = \frac{1}{\abs{E_G^\epsilon}}\sum_{e \in E_G^\epsilon} \frac{1}{\abs{B_r\pof{e}}}\sum_{\substack{k \\ v_k \in B_r\pof{e}}} \pof{\kappa_e^{\text{R}} - \kappa^{\text{G}}_k}^2.\]

\paragraph{Reverse Computation}
Differentiating, \begin{alignat*}{1}
    &\frac{\partial\pof{\mathcal{L}_{\mathrm{curvature}}\pof{M}}}{\partial z_\ell} \\
    &\qquad= \frac{1}{\abs{E_G}}\sum_{e \in E_G} \frac{1}{\abs{B_r\pof{e}}}\sum_{\substack{k \\ v_k \in B_r\pof{e}}} -2\pof{\kappa_e^{\text{R}} - \kappa^G_k}\frac{\partial\kappa^G_k}{\partial z_\ell}.
\end{alignat*}

\subsubsection{Smoothness Loss}
\paragraph{Forward Computation}
We use \[\mathcal{L}_{\mathrm{smooth}}\pof{M} = \pof{-\pof{\kappa^+}^\intercal L_C^{\text{D}}\kappa^+ - \pof{\kappa^-}^\intercal L_C^{\text{D}}\kappa^-} \sum_{\pof{v_i, v_j} \in E_M} A_{i, j}.\]

\paragraph{Reverse Computation}
Differentiating, and using matrix symmetry, \begin{alignat*}{1}
    &\frac{\partial\pof{\mathcal{L}_{\mathrm{smooth}}\pof{M}}}{\partial z_\ell} \\
    &\qquad= -\pof{\kappa^+}^\intercal\pof{\frac{\partial L_C^{\text{D}}}{\partial z_\ell}\kappa^+ + 2L_C^{\text{D}}\frac{\partial\kappa^+}{\partial z_\ell}} \sum_{\pof{v_i, v_j} \in E_M} A_{i, j} \\
    &\qquad\qquad- \pof{\kappa^-}^\intercal\pof{\frac{\partial L_C^{\text{D}}}{\partial z_\ell}\kappa^- + 2L_C^{\text{D}}\frac{\partial\kappa^-}{\partial z_\ell}} \sum_{\pof{v_i, v_j} \in E_M} A_{i, j} \\
    &\qquad\qquad- \pof{\pof{\kappa^+}^\intercal L_C^{\text{D}}\kappa^+ + \pof{\kappa^-}^\intercal L_C^{\text{D}}\kappa^-} \sum_{\pof{v_i, v_j} \in E_M} \frac{\partial A_{i, j}}{\partial z_\ell}.
\end{alignat*}

\subsection{Visualization Tool Implementation Details}

\subsubsection{Data Flow}
\begin{figure}[t]
    \centering
    \includegraphics[width=\columnwidth]{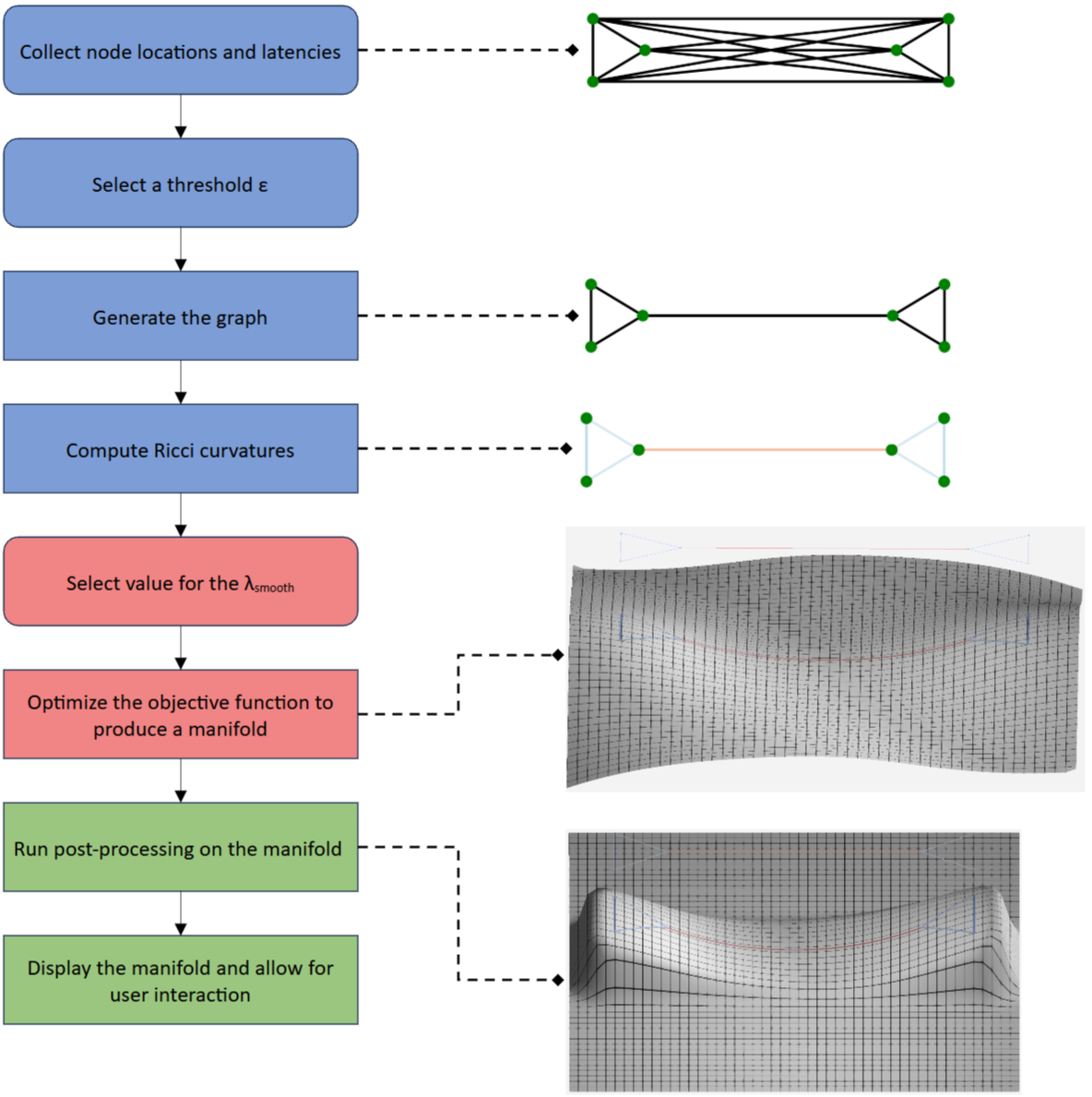}
    \caption{Flowchart of how the Matisse application is run. Blue boxes compute the network, red boxes compute the manifold, and green boxes display the user interface. Boxes with rounded corners require user input to collect data and set parameters. Images to the right show the graph and/or manifold as it has been computed at that step.}
    \label{fig:flowchart}
\end{figure}

\begin{figure}[t]
    \centering
    \includegraphics[width=\columnwidth]{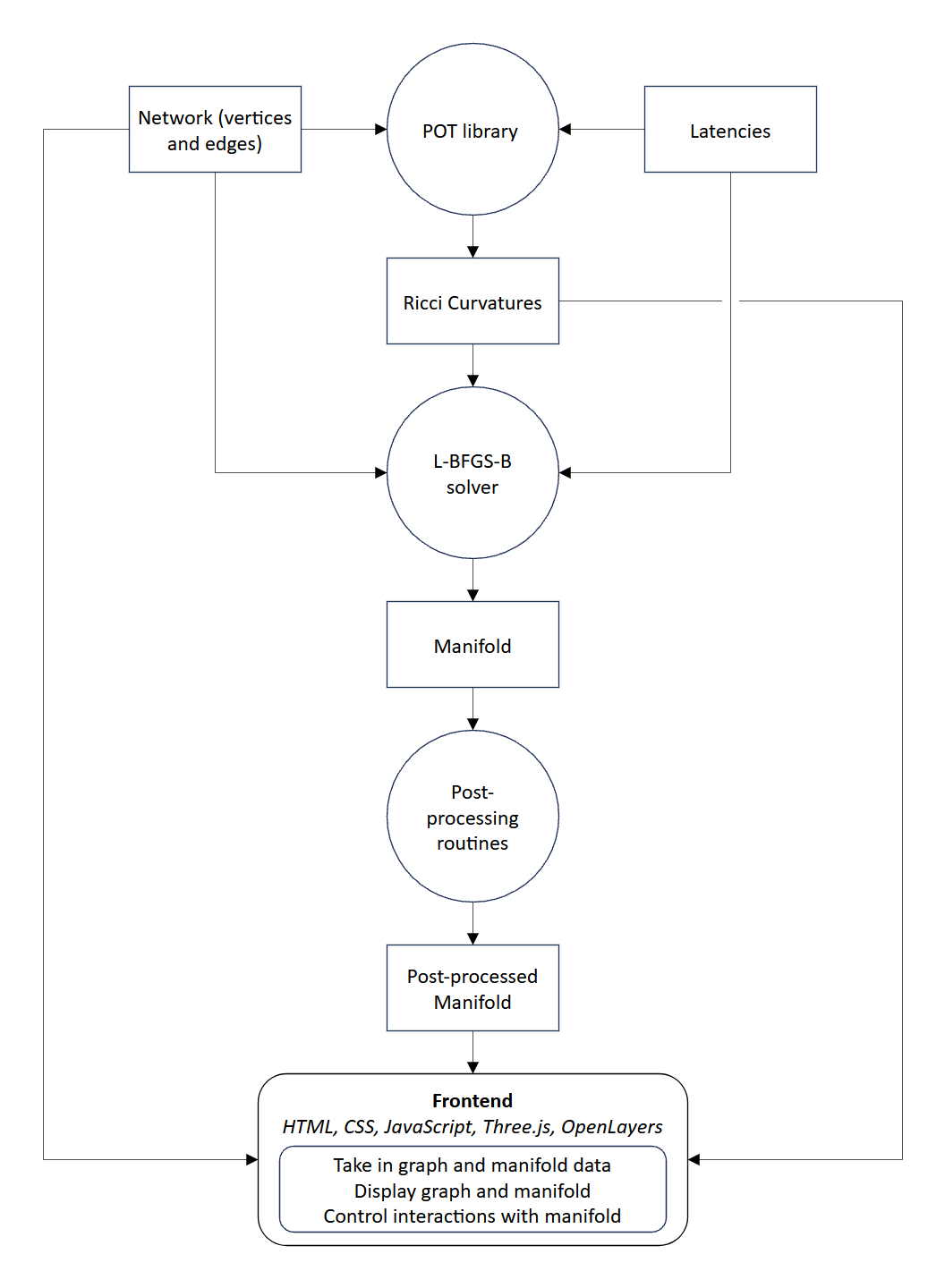}
    \caption{Data flow of the Matisse application. Boxes with square corners represent data, circles represent functions that are run offline, and boxes with rounded corners represent functions that are run online. The arrows show the flow of the data.}
    \label{fig:dataflow}
 \end{figure}

\Cref{fig:flowchart} describes the pipeline for Matisse from user input to visualization output, and a diagram depicting the data flow of Matisse is provided in \cref{fig:dataflow}.

\subsubsection{Performance}
The outputs in this paper were obtained on an AMD Ryzen 7 3700X processor. For one particular run of the optimization algorithm on the US dataset with delay threshold $10$ on a $50 \times 50$ mesh, each iteration lasted an average of $1.85$ seconds. The algorithm converged after $14698$ iterations, resulting in a total runtime of seven and a half hours. While this runtime is large, it comes with a couple of caveats. First, the early stopping of the optimization allows for cutting down the runtime at the expense of a less accurate output. Limiting to just $2000$ iterations yielded comparable output in practice. Second, the system described in this paper is thus written mostly in Python. Rewriting the computations in C or another compiled language would no doubt provide a significant speedup, and will be done in future work.

Finally, from \cref{sec:details_of_computation}, we can place some theoretical bounds on the runtime of each evaluation of the loss function and its gradient. We find that each evaluation scales roughly as $O(n + |E|\sqrt{n})$, where $n$ is the number of vertices in our square mesh.

\subsection{Additional Real World Applications}

\subsubsection{Additional Views of the Measured Delay Space for the US Public Internet}
\label{sec:internet_US_extra}
\begin{figure}[ht]
    \centering
    \includegraphics[width=\columnwidth]{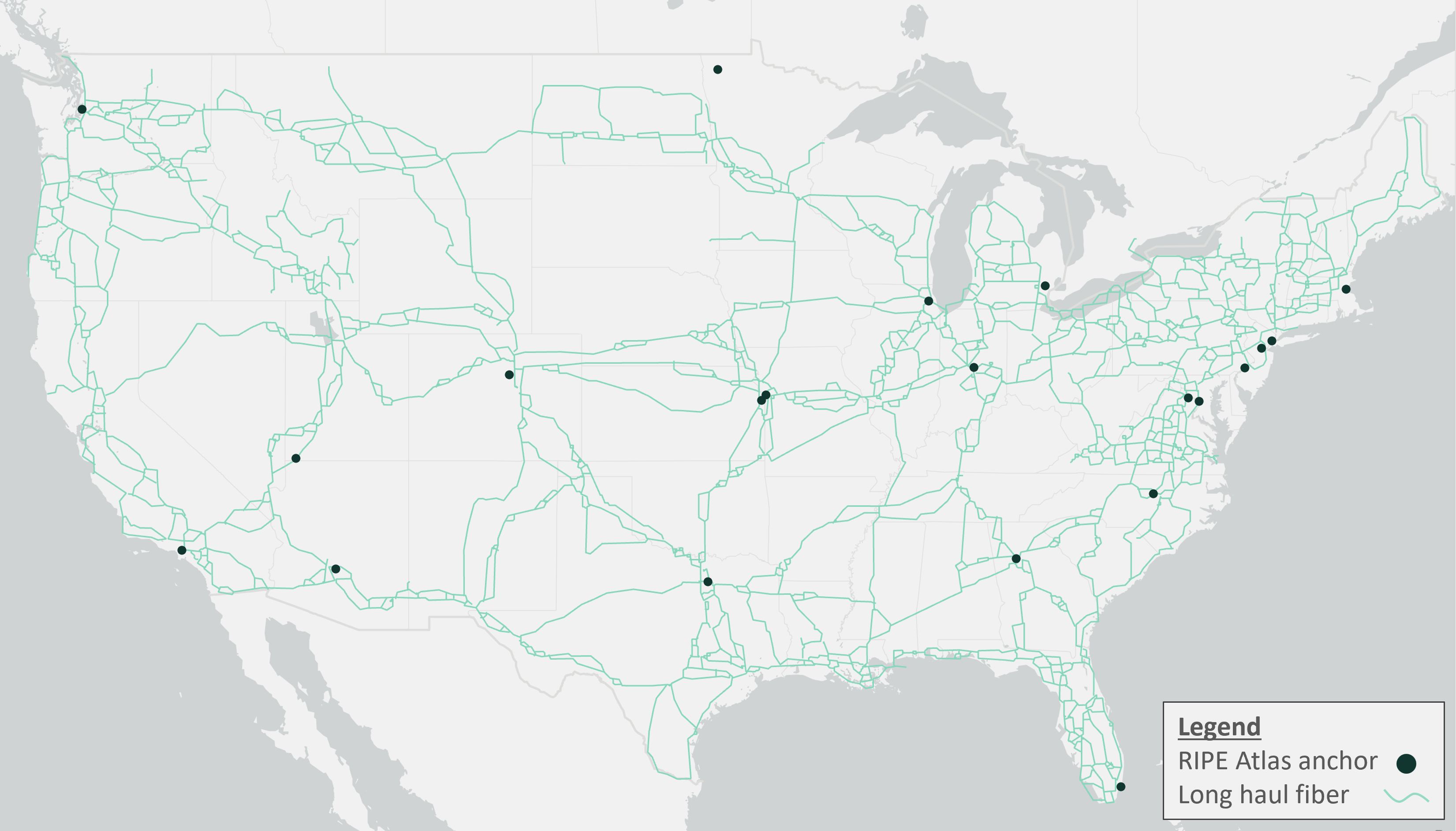}
    \caption{Geographic map of the RIPE Atlas anchors and long haul fiber network in the US.}
    \label{fig:us_atlas_map}
\end{figure}

We provide a reference map of the the RIPE Atlas anchors in the US in \cref{fig:us_atlas_map}.

\begin{figure}[ht]
    \centering
    \begin{subfigure}{\columnwidth}
        \centering
        \includegraphics[width=0.8\columnwidth]{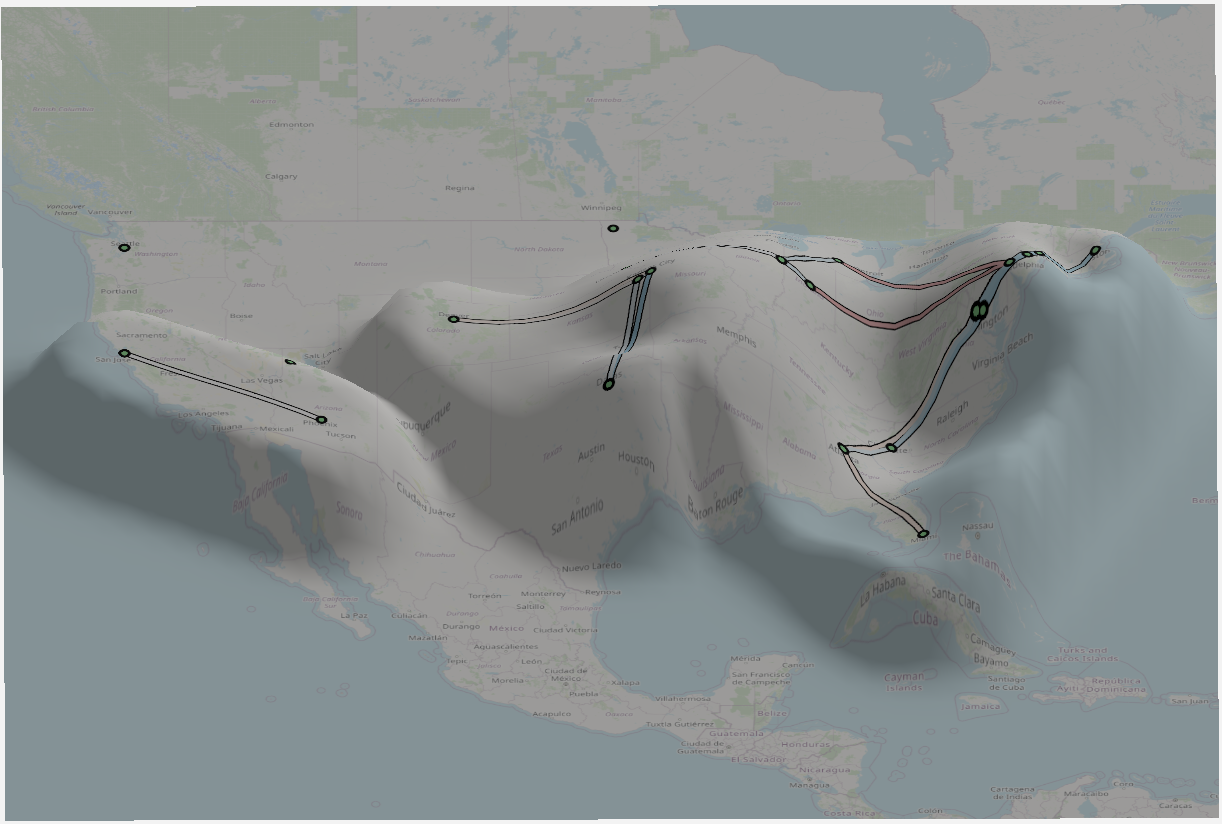}
        \caption{The US Internet delay space at low residual delay threshold.}
        \label{fig:public_us_extra_low}
    \end{subfigure}
    \begin{subfigure}{\columnwidth}
        \centering
        \includegraphics[width=0.8\columnwidth]{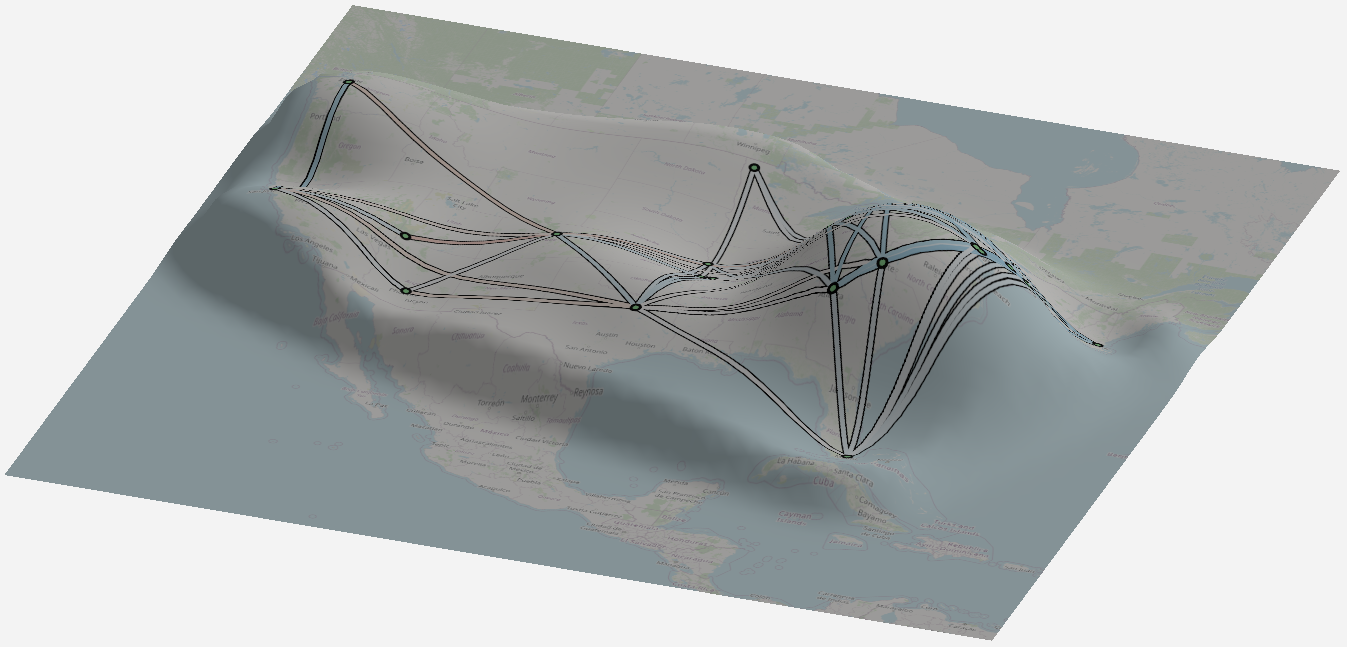}
        \caption{The US Internet delay space at a high residual delay threshold.}
        \label{fig:public_us_extra_high}
    \end{subfigure}
    \begin{subfigure}{\columnwidth}
        \centering
        \includegraphics[width=0.8\columnwidth]{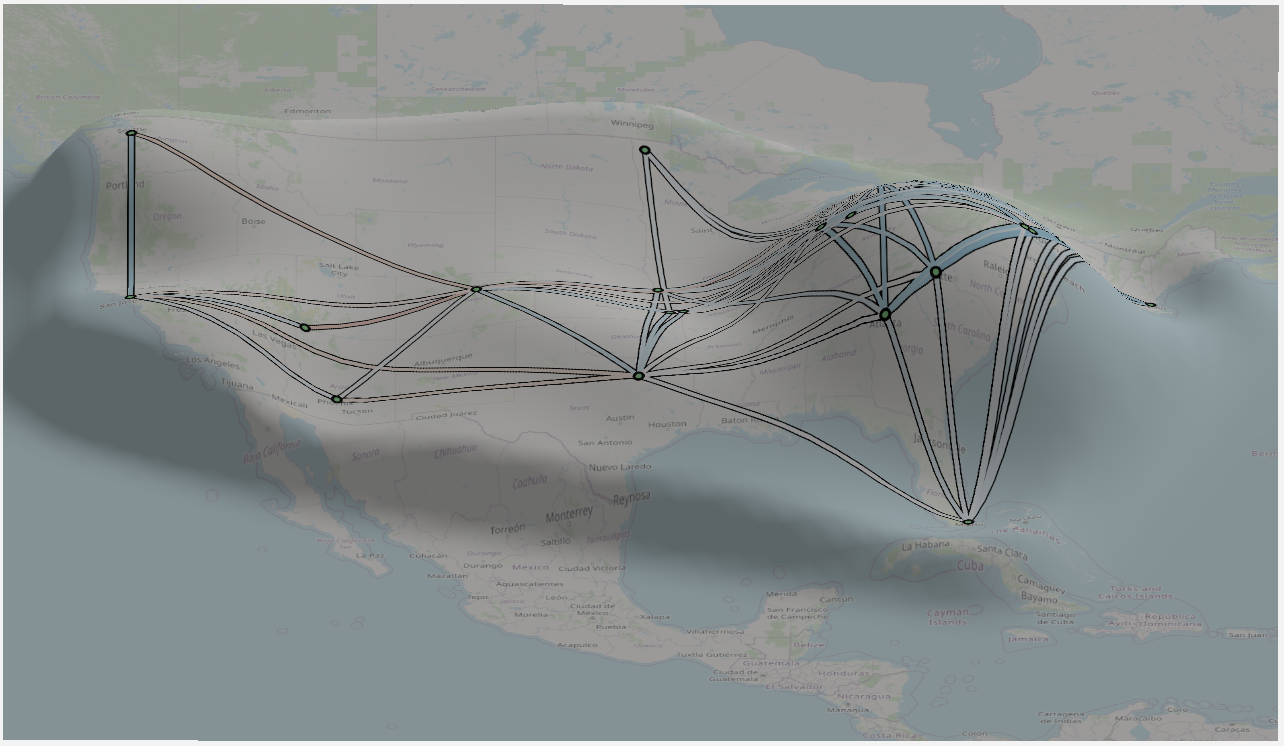}
        \caption{Alternative view of the US Internet delay space at a high residual delay threshold.}
        \label{fig:public_us_extra_high_alt}
    \end{subfigure}
    \caption{Manifold representations of the US public Internet at different residual delay thresholds.}
    \label{fig:public_us_extra}
\end{figure}

One major benefit for Matisse as a tool is its interactive abilities. By panning and rotating the camera, the user can avoid issues caused by occlusion. Due to the nature of a document, we cannot fully show off Matisse's capabilities here. We do, however, provide some possible alternative views of the US examples from \cref{fig:public_us} in \cref{fig:public_us_extra}.

\Cref{fig:public_us_extra_high,fig:public_us_extra_high_alt} shows the manifolds for connectivity that at a higher residual threshold ({\em i.e.,} including links with residual delays further away from GCL). These figures highlight a more nuanced picture of the U.S. Internet than \cref{fig:public_us,fig:public_us_extra_low}, one that retains regional distinctions within a singularly connected national framework. In particular, we identify clear saddles in geographic areas such as the Rocky Mountains and parts of the Southern U.S. ({\em e.g.,} the Mississippi Delta) where the topography of the land precludes the deployment of extensive physical infrastructure and limits fiber-optic routes to a small number of conduits, typically along the few existing interstate highways or railroad tracks.

\subsubsection{Graphical Comparison Between RTTs and Geodesic Distances}
\begin{figure}[ht]
    \centering
    \includegraphics[width=\columnwidth]{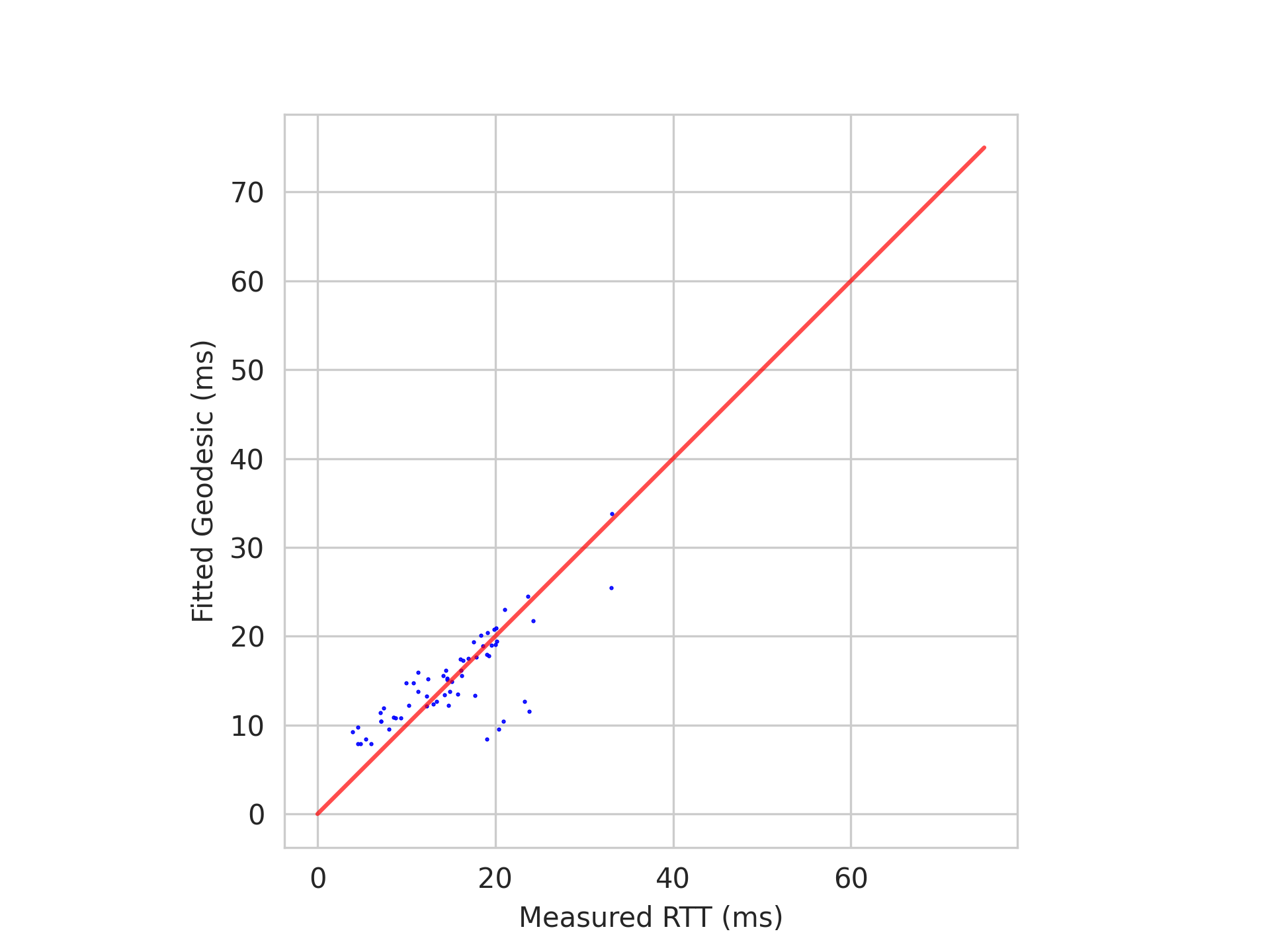}
    \caption{Fitted geodesics against measured latencies for the US data with delay threshold $10$ milliseconds.}
    \label{fig:latency_scatter}
\end{figure}

\begin{figure}[ht]
    \centering
    \includegraphics[width=\columnwidth]{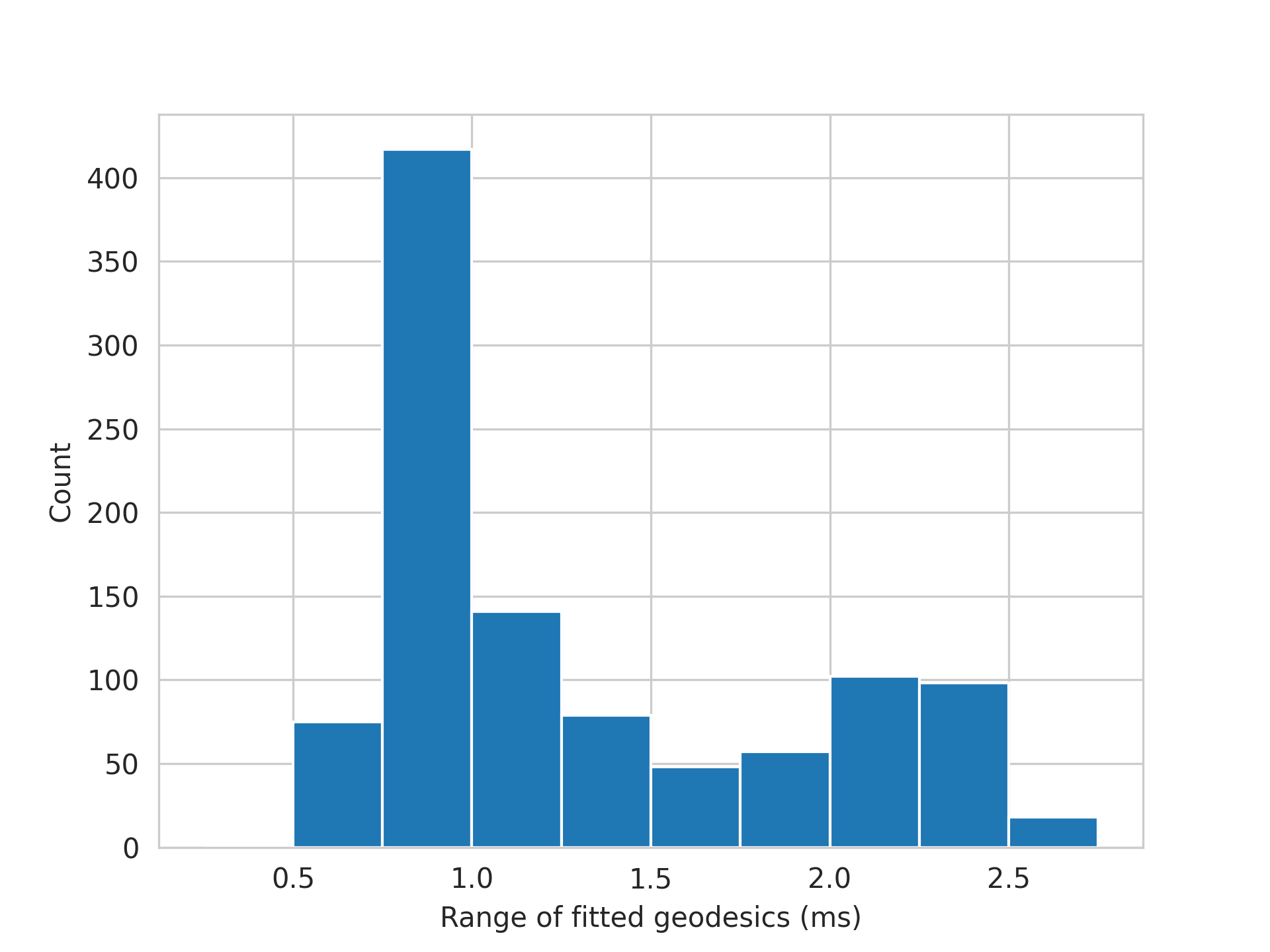}
    \caption{Ranges of fitted geodesics in milliseconds for fixed city pairs across two weeks. Data are from the US dataset with delay threshold $22$ milliseconds.}
    \label{fig:latency_histogram}
\end{figure}

We further explore the results from \cref{sec:manifold_distance}. \Cref{fig:latency_scatter} shows that the geodesic distances derived from our manifold generally reflect the measured RTTs. \Cref{fig:latency_histogram} displays the stability of the geodesic distances across a time span of weeks.

\label{sec:graphical_delay_space}

\subsubsection{Measured Delay Space for the European Public Internet}
\label{sec:internet_Europe}
\begin{figure}[p]
   \centering
    \begin{subfigure}{\columnwidth}
        \centering
        \includegraphics[width=0.8\columnwidth]{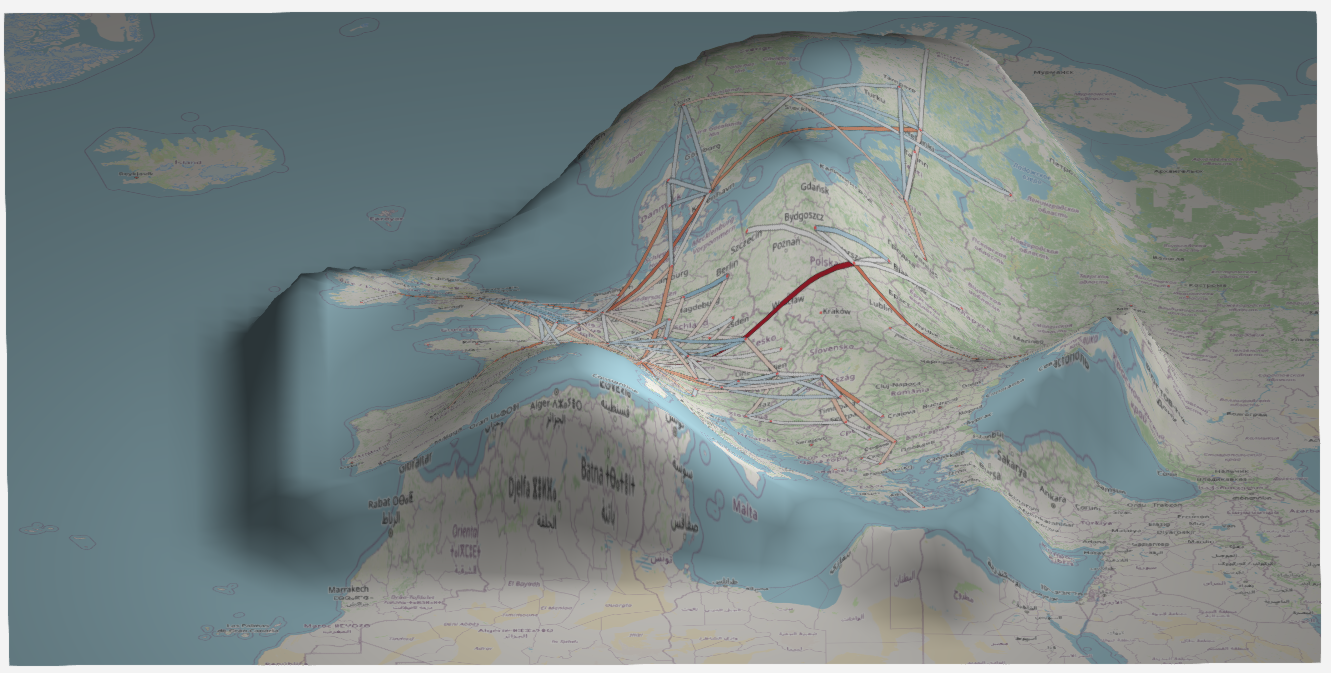}
        \caption{European Internet delay space.}
        \label{fig:manifold_Europe_clustered_front}
    \end{subfigure}
   \begin{subfigure}{\columnwidth}
       \centering
       \includegraphics[width=0.8\columnwidth]{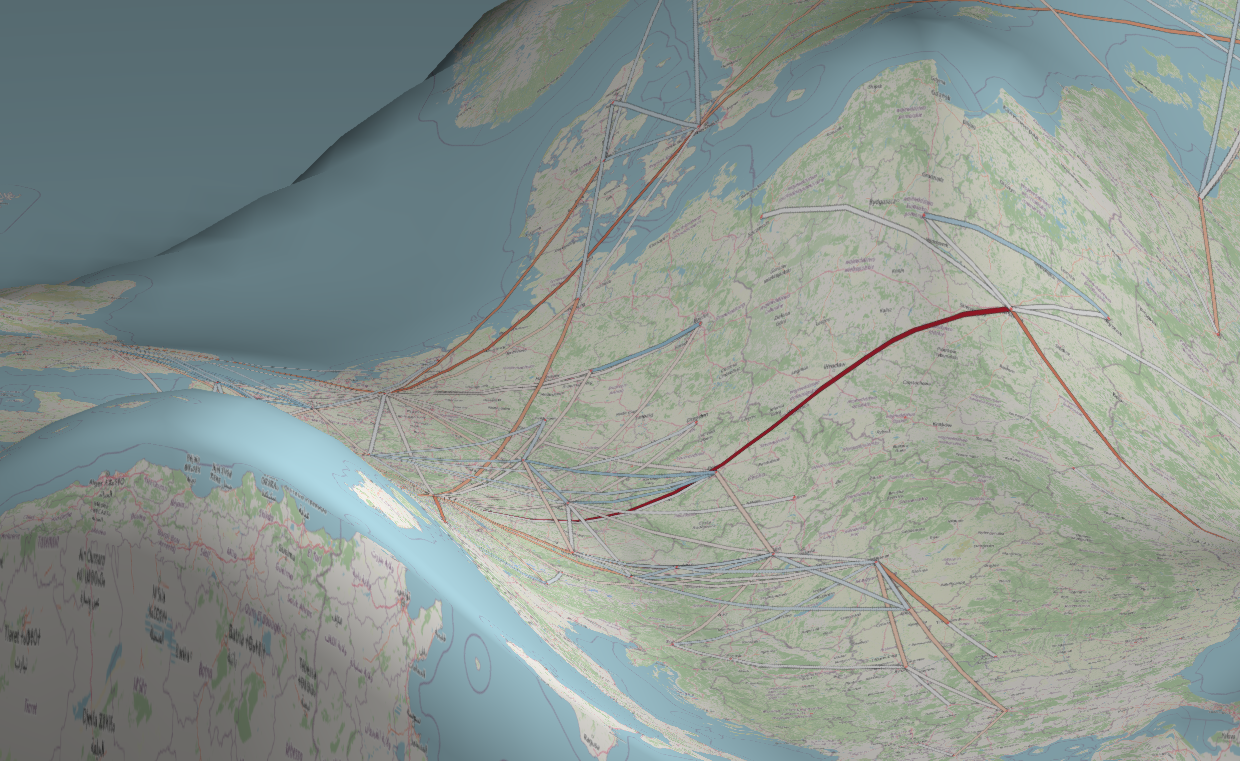}
       \caption{Zoomed section of European Internet delay space.}
       \label{fig:manifold_Europe_clustered_zoomed}
   \end{subfigure}
   \caption{Manifold representations of the European public Internet at a medium residual delay threshold (5 ms), using a clustered input network.}
   \label{fig:manifold_Europe_clustered}
\end{figure}

In another case study, we consider the European public Internet. Similar to our case study of the US public Internet, we use data from RIPE Atlas to generate a connectivity graph and resulting manifolds via Matisse. \Cref{fig:manifold_Europe_clustered} shows both zoomed out and zoomed in manifolds for a medium residual delay threshold (5 ms), using a clustered input network.  At this threshold, we can see small connectivity islands within metropolitan areas as well as negatively curved edges, which serve as connectivity bridges between these metropolitan areas. These edges intuitively represent physical conduits linking different clusters of connection and represent ``important'' links in the network.

\begin{figure}[p]
    \centering
    \begin{subfigure}{\columnwidth}
        \centering
        \includegraphics[width=\columnwidth]{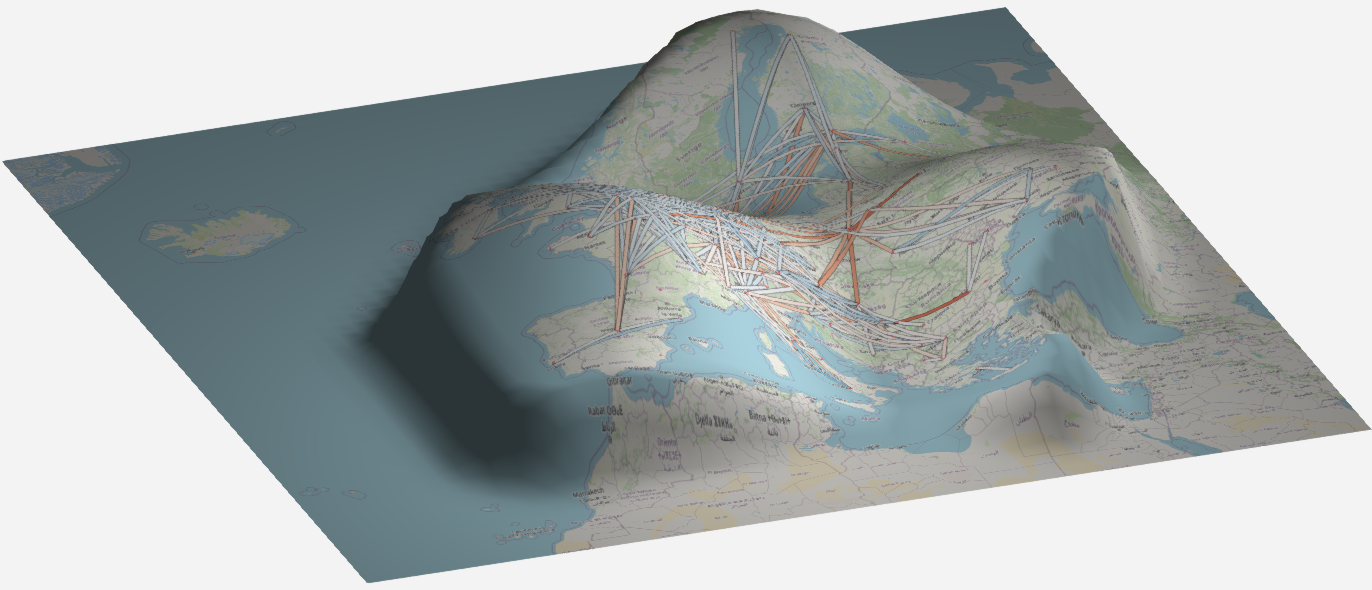}
        \caption{Europe Internet delay space.}
    \end{subfigure}
    \begin{subfigure}{\columnwidth}
        \centering
        \includegraphics[width=\columnwidth]{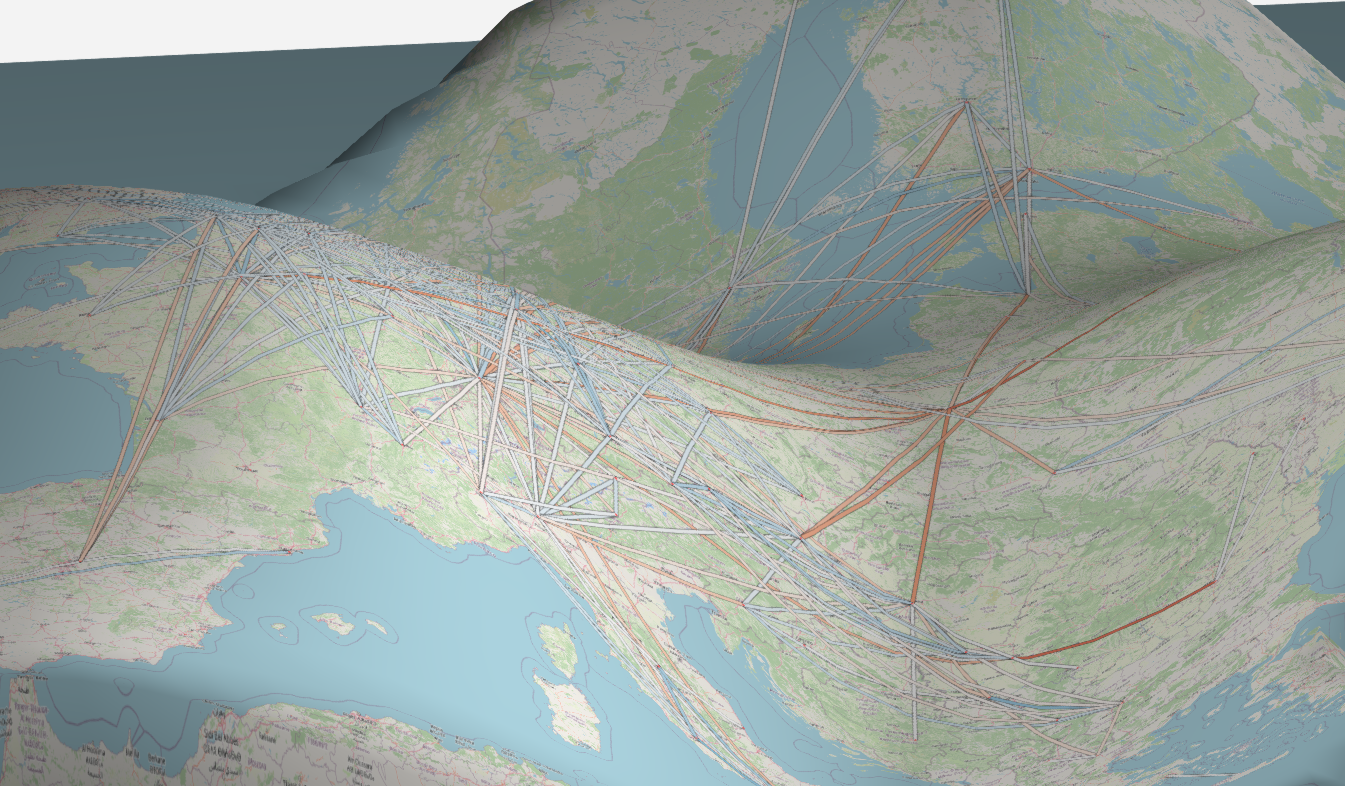}
        \caption{Zoomed section of Europe Internet delay space.}
    \end{subfigure}
    \caption{Manifold representations of the European public Internet at a high residual latency threshold ($8$ ms), using a clustered input network.}
    \label{fig:manifold_Europe_extra}
\end{figure}

\Cref{fig:manifold_Europe_extra} shows both  zoomed out and zoomed in visualizations of the European public Internet generated with a relatively high residual latency threshold ($8$ ms) and using a clustered input network.  As a result of the threshold, the resulting graph is relatively dense in terms of link connectivity.  The manifold visualization shows how densely connected the regions corresponding to Western Europe and Scandinavia are, as exhibited by the large hills reflecting positive Ricci curvature in the graph. The visualization also highlights the relative lack of connectivity through Eastern Europe, as evidenced by the saddle feature in that region.  The diverse visuals in \cref{fig:manifold_Europe_clustered,fig:manifold_Europe_extra} emphasize different aspects of connectivity across Europe. Matisse was specifically designed to facilitate the study and analysis of these various levels of granularity, as is explained in  \cite{sigmetrics2022}.

\end{document}